\documentclass[%
 aip,
 amsmath,amssymb,
reprint,%
]{revtex4-2}

\usepackage{graphicx}
\usepackage{dcolumn}
\usepackage{bm}

\usepackage[utf8]{inputenc}
\usepackage[T1]{fontenc}
\usepackage{mathptmx}
\usepackage{etoolbox}
\usepackage{comment}

\usepackage{physics} 
\usepackage{amssymb} 

\newcommand{\ford}[1]{#1\textsuperscript{(1)}}
\newcommand{\aux}[1]{#1\textsubscript{aux}}

\newcommand{\bl}[1]{\mathbf{#1}}
\newcommand{\bsigma}{\boldsymbol{\sigma}}
\newcommand{\btau}{\boldsymbol{\tau}}
\newcommand{\avg}[1]{\langle #1 \rangle}

\newcommand{\snabla}{\boldsymbol{\nabla}_s}
\newcommand{\bnabla}{\boldsymbol{\nabla}}

\newcommand{\bfu}{\mathbf{u}} 
\newcommand{\bfx}{\mathbf{x}} 
\newcommand{\bfk}{\mathbf{k}} 
\newcommand{\bcdot}{\boldsymbol{\cdot}}

\usepackage{color}

\makeatletter
\def\@email#1#2{%
 \endgroup
 \patchcmd{\titleblock@produce}
  {\frontmatter@RRAPformat}
  {\frontmatter@RRAPformat{\produce@RRAP{*#1\href{mailto:#2}{#2}}}\frontmatter@RRAPformat}
  {}{}
}%
\makeatother
\begin{document}

\preprint{AIP/123-QED}

\title{Hydrodynamic Aggregation of Membrane Inclusions due to Non-Newtonian Surface Rheology} 
\author{Vishnu Vig}
\author{Harishankar Manikantan}
\affiliation{ 
Department of Chemical Engineering, UC Davis, Davis, CA
}%
 \email{hmanikantan@ucdavis.edu}%

\date{\today}

\begin{abstract}
Biological membranes are self-assembled complex fluid interfaces that host proteins, molecular motors and other macromolecules essential for cellular function. These membranes have a distinct in-plane fluid response with a surface viscosity that has been well characterized. The resulting quasi-2D fluid dynamical problem describes the motion of embedded proteins or particles. However, the viscous response of biological membranes is often non-Newtonian: in particular, the surface shear viscosity of phospholipids that comprise the membrane depends strongly on the surface pressure. We use the Lorentz reciprocal theorem to extract the effective long-ranged hydrodynamic interaction among membrane inclusions that arises due to such non-trivial rheology. We show that the corrective force that emerges ties back to the interplay between membrane flow and non-constant viscosity, which suggests a mechanism for biologically favorable protein aggregation within membranes. We quantify and describe the mechanism for such a large-scale concentration instability using a mean-field model. Finally, we employ numerical simulations to demonstrate the formation of hexatic crystals due to the effective hydrodynamic interactions within the membrane.
\end{abstract}

\maketitle


\section{Introduction}

Lipids and other surface-active macromolecules play critical roles in biological interfaces. Membranes of eukaryotic cells are self-assembled phospholipid bilayers \citep{rob,lip}. These complex systems display rich dynamics due to their inherent quasi-2D viscous nature that emerges from the coupling to the surrounding bulk phases \cite{Saffman1975,Hughes1981}. Structural and rheological properties of fluid membranes have been well characterized in the past decades, conceptually driven by connecting in-plane protein diffusion to membrane fluidity. Surface viscosities are now known for single and multi-component lipid monolayers \citep{Choi2011,Espinosa2011,Samaniuk2014}, phospholipid bilayers \citep{Gambin2006,Cicuta2007,Schutte2017}, protein-laden interfaces \citep{Prasad2006}, and polymeric multilayers \citep{Pepicelli2019}. Simultaneously, continuum fluid dynamical models have been widely successful in describing the isolated and cooperative motion of disk-like \citep{Stone1998,Oppenheimer2009,Camley2013}, rod-like \cite{Fischer2004,Levine2004,Shi2022}, active \cite{actprot,Manikantan2020PRL,Oppenheimer2019}, and polymeric \cite{Ramachandran2011} inclusions in simple 2D Newtonian membranes.

However, the rheology of real membranes is rarely Newtonian. Insoluble surfactant species often phase separate to form 2D dispersions or `rafts' of condensed liquid crystalline phases suspended in a liquid expanded disordered phase \cite{Kaganer1999,Lingwood2010,Sezgin2017}. Unlike 3D fluids, surfactants are much easier to compress and even a facile in-plane deformation triggers significant changes in the rheological signature of the lipid interface. The transport of surfactant molecules between coexisting phases as well as the morphology and interlocking of stiffer domains contribute to a strongly nonlinear surface rheological response. Insoluble surfactant layers exhibit jamming and yield stress behavior \citep{Choi2011}, surface viscoelasticity \citep{Arriaga2010}, surface shear thinning \citep{Kurnaz1997}, and surface-concentration-dependent surface rheology \citep{Kim2011,Samaniuk2014}.  While a lot of the intuition of 3D fluid mechanics can be ported to the analysis of 2D viscous manifolds, the nature of momentum transport in interfacial systems and the novel rheological relations present unique challenges. 

In this work, we aim to theoretically investigate and numerically demonstrate the role of a class of non-Newtonian membrane behavior on driven or anchored trans-membrane proteins or protein-associated domains that play critical roles in cell motility and structure \cite{Fogelson2014,Bruinsma1996}. These membrane inclusions are anchored to and driven relative to the membrane by the intracellular network of actin and microtubules, or by external motor proteins such as kinesin or dynein \cite{rob}. We will specifically target the nontrivial hydrodynamic interactions expected to arise due to surface-pressure--dependent surface viscosity of lipids. Despite this strong nonlinearity, recent mathematical efforts have demonstrated the qualitatively new phenomena that emerge in thin interfacial gaps \citep{HMPRF}, in pore-spanning monolayers or membranes \citep{HMPRSA}, and in the resulting stability and dynamics of drops containing such surfactants \citep{Singh2021,Herrada2022}. These past works point to kinematic symmetry breaking within the 2D interfacial layer, leading to trajectories of probe particles, lipid rafts, or naturally occuring membrane inclusions that are not expected in Newtonian systems. Yet, the exact nature of the hydrodynamic interactions between pairs of driven inclusions representing cytoskeletal anchors, membrane motors, and adhesion junction proteins remain a mystery. We derive analytic results for such hydrodynamic interactions in this work, shedding light on the effective inter-particle attraction or repulsion between interfacially driven particles. Collectively, such interactions can lead to large-scale aggregation, which might be biologically favorable in immune response, locomotion, and tubulation \cite{Bruinsma1996}. Past work on membrane inclusions have demonstrated long-ranged interactions and assembly purely due to protein activity \citep{Oppenheimer2019,Manikantan2020PRL}, membrane curvature \citep{Reynwar2007,Dasgupta2017}, and inclusion size mismatch \cite{Bruinsma1996}. Building on our analysis of pair dynamics, we demonstrate here that non-Newtonian surface rheological response can also lead to aggregation and packing of `2D suspensions' of driven or anchored inclusions on the membrane. 

This paper is organized as follow: we first evaluate the motion of a single particle driven in a rheologically complex membrane in \S \ref{sec:single}. In this section, we formulate the fluid dynamical problem of a disk-like particle embedded in the membrane and driven by an external force while subject to the 2D flow in the plane of the membrane. We will develop a perturbative solution for weakly non-Newtonian behavior, derive the corrected velocity of the particle, and physically rationalize the result. In \S  \ref{sec:collective}, we extend the study to collective behavior via membrane hydrodynamic interactions due to `2D suspensions' of particles driven in the plane of a membrane. We propose a mean-field description to quantify the instability that leads to large-scale aggregation, develop a Langevin description that allows us to simulate multi-particle dynamics, and explore the hexatic order that emerges from crystalline aggregation due to attractive hydrodynamic interactions. We conclude with a discussion of the relevance of these results to real biological systems, and connections to other problems where viscosity depends on pressure.

\section{Single particle dynamics in non-Newtonian membranes}\label{sec:single}

\subsection{Problem formulation}

\begin{figure}[]
	\includegraphics[width=\linewidth]{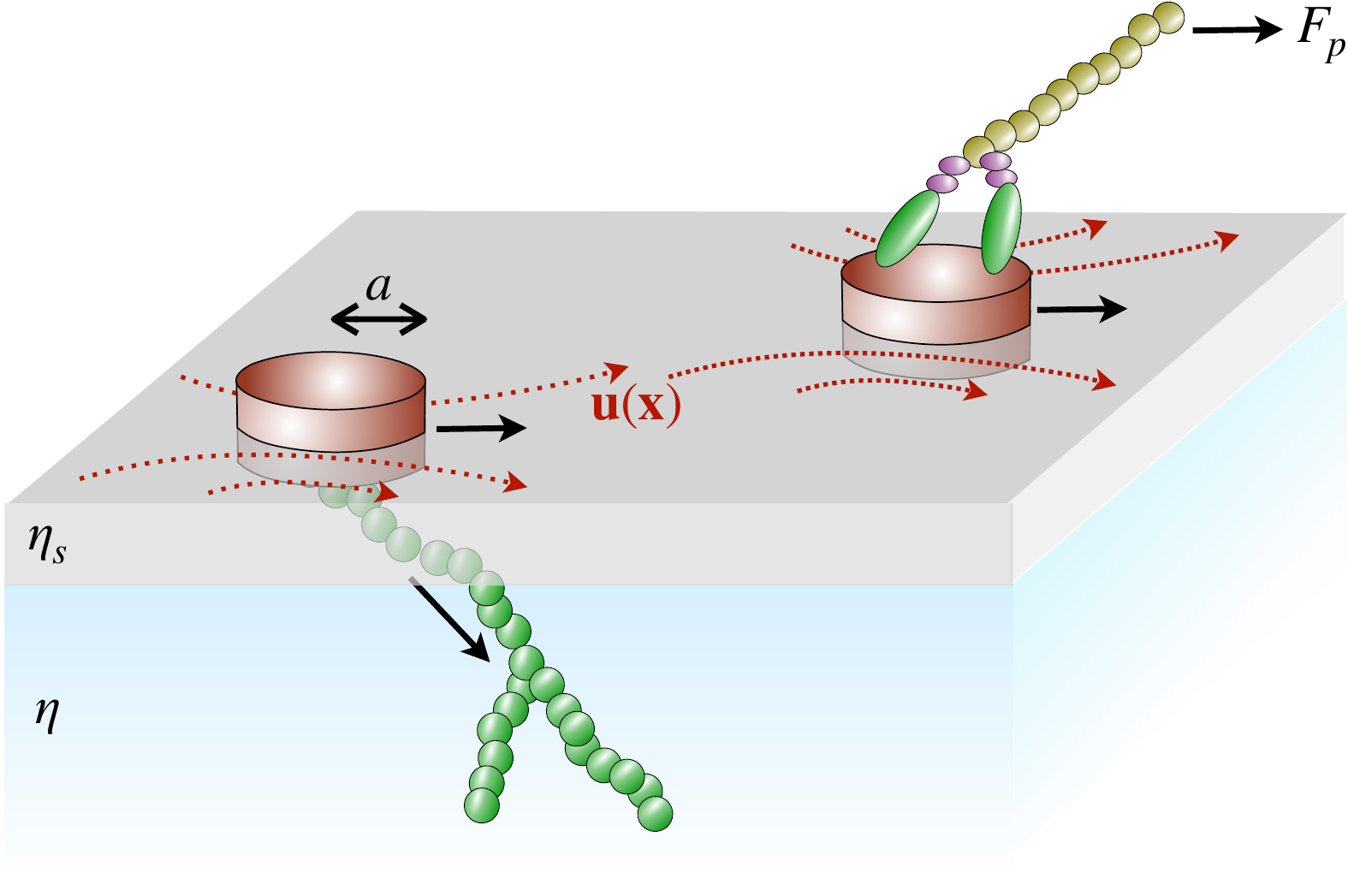}
	\caption{Illustration of a disk-like inclusion representing a transmembrane protein embedded in a phospholipid membrane. The inclusion may be driven by intracellular forces exerted by the cytoskeleton, or by extracellular forces due to motor protein activity during transport or at adhesion junctions. The fluid dynamical problem is equivalent to an anchored immobile inclusion within a flowing membrane. Forces driving such inclusions can generate in-plane hydrodynamic disturbance flows and interactions.}
	\label{fig:intro}
\end{figure}

The geometry of our fluid dynamical system is shown in Fig.~\ref{fig:intro}. A bulk fluid phase of viscosity $\eta$ representing intracellular space underlies a membrane of potentially non-constant surface shear viscosity $\eta_s$. Fluid flow in the bulk phase with velocity $\bl{v}$ is governed by the Navier-Stokes equation, with a no-slip condition to match the membrane velocity $\bfu$. Additionally, the stress jump at the interface describes conservation of 2D momentum \citep{Slate,Manikantan2020}:
\begin{equation}
	\rho_s \frac{D\bfu_s}{Dt} = \snabla \bcdot \bsigma_s - \eta \pdv{\bl{v}}{z} \bigg|_{\text{z}=0},
	\label{cauchy}
\end{equation}
where we have assumed a flat membrane along the $x$--$y$ plane for simplicity, and $D/Dt$ is the material derivative. $\snabla = \bl{I}_s \bcdot \bnabla$ is the surface gradient operator with $\bl{I}_s = \bl{I} - \hat{\bl{n}} \hat{\bl{n}} $ the surface identity tensor on a plane with unit normal $\hat{\bl{n}}$. Equation~\eqref{cauchy} is readily generalized to bulk fluids on either side of the membrane: the results that follow are qualitatively unchanged and are only modified by a prefactor in that case.

Equation~\eqref{cauchy} is a 2D Cauchy equation for a species of density $\rho_s$ and 2D stress tensor $\bsigma_s$ that is forced via viscous tractions from the adjacent bulk phase(s). Alternatively, we may interpret Eq.~\eqref{cauchy} as a boundary condition for the Navier-Stokes equation governing 3D fluid field $\bl{v}$. The 2D stress tensor $\bsigma_s$ may be decomposed into isotropic and deviatoric parts:
\begin{equation} 
	\bsigma_s = - \Pi \, \bl{I}_s + \btau_s,
\end{equation}
where $\Pi$ is the 2D pressure and the deviatoric stress tensor $\btau_s$ is prescribed by a constitutive relation. For instance, 
\begin{equation}
\btau_s=  \eta_s \left[ \snabla \bfu \bcdot \bl{I}_s  + \bl{I}_s \bcdot (\snabla \bfu)^{\text{T}}  \right]
\end{equation}
in the simple case of a 2D Newtonian membrane. Underlying such a constitutive relation is the approximation that the interface is 2D incompressible. In other words, we assume that the surface flow rearranges to that of a 2D incompressible material so long as Marangoni flows (driven by gradients in the surface pressure $\Pi$) are generated faster than the rate of in-plane compression. This is almost always the case for insoluble phospholipids that make up most biological membranes  \citep{Manikantan2020,Fischer_comment,Stone2015}. Then, neglecting inertia in these overdamped systems, Eq.~\eqref{cauchy} simplifies to
\begin{equation}
	\label{goveq}
	\eta \pdv{\bl{v}}{z} \bigg|_{\text{z}=0} = -\snabla \Pi + \eta_s\nabla^2\bfu_s, \quad \snabla\bcdot\bfu_s = 0.
\end{equation}   
Equation~\eqref{goveq} governs momentum and mass conservation of an insoluble incompressible Newtonian membrane with constant surface shear viscosity. A wide range of theoretical studies spanning decades \citep{Saffman1975,Hughes1981,Stone1998,Oppenheimer2009} describing in-plane motion of particles in Newtonian membranes start with Eq.~\eqref{goveq}.

However, we intend to explore the role of non-constant $\eta_s$ on membrane inclusion dynamics. Specifically, we wish to account for surface-pressure-dependent surface viscosity of lipids. In what follows, we assume that surface pressure enforces incompressibility and $\Pi$-dependent viscosities are incorporated by treating the the interfacial momentum equation \eqref{goveq} as a generalized Newtonian model. This is analogous to dropping the equation of state for density in the incompressible Navier-Stokes equation: the thermodynamic quantity $\Pi$ plays the role of the `mechanical' pressure within this approximation. The validity of such an approximation is diskussed in detail in past works \citep{HMPRF} and this model has been widely successful in studying these complex systems \citep{HMPRSA,Singh2021,Herrada2022}.  

Such a description of the non-constant surface shear viscosity also relates directly to typical experiments where the surface viscosity is measured as a function of surface pressure. $\eta_s$ changes order of magnitude over a range of $\Pi$ commonly accessible in experiments, unlike 3D fluids whose viscosity only changes under extreme pressures, if at all. For example, the phospholipid dipalmitoylphosphatidylcholine (DPPC) is a major constituent of cell membranes and forms stable monolayers at an air-water interface that undergo a liquid expanded (LE) to liquid condensed (LC) phase transition at $\Pi \sim 8 \,\text{mN/ m}$ at room temperature \citep{Kaganer1999}. Above this critical surface-pressure, the monolayer viscosity increases exponentially with surface pressure \citep{Kim2011}. Similar `$\Pi$-thickening' behavior has been observed in certain fatty acids like nonadaconic (C$_{19}$) acid, heneicosanoic (C$_{21}$) acid and behenic (C$_{22}$) acid \citep{Alonso2006}. Conversely, phase transitions associated with chain-tilt lead to molecules shearing past each other: as a result $\eta_s$ decreases exponentially with increasing $\Pi$ from $10$ to $20\, \text{mN/m}$ in the `$\Pi$-thinning' surfactant eicosanol \citep{Zell2014}.

The simplest constitutive relation for $\eta_s(\Pi)$ follows free-area models of surface viscosity \citep{Alonso2006,Kim2013} to give an exponential relation:
\begin{equation}\label{eq:etapi}
	\eta_s (\Pi)  = \eta^0_{s} e^{\Pi-\Pi_\infty/\Pi_c},
\end{equation}
where  $\Pi_c$ is the characteristic surface pressure change required to produce a noticeable change in $\eta_s$, and $\eta^0_{s}$ is a reference or unperturbed viscosity at a reference pressure $\Pi_\infty$. Setting $\Pi_c \rightarrow \infty$ retrieves the Newtonian limit of constant surface viscosity. Equation~\eqref{eq:etapi} accommodates both $\Pi$-thickening ($\Pi_c>0$) and  $\Pi$-thinning ($\Pi_c<0$) surfactants. Plugging Eq.~\eqref{eq:etapi} into the Cauchy momentum equation and simplifying in the overdamped limit then gives
\begin{equation}
-{\bnabla}_s  \Pi + {\bnabla}_s \bcdot \left[  \eta_s( \Pi) \left({\bnabla}_s \bl{ u}_s +{\bnabla}_s \bl{ u}_s^T \right)\right]= \eta \pdv{\bl{ v}}{ z} \bigg|_{\text{z}=0},
\end{equation}
which along with the incompressibility condition $\snabla\bcdot\bfu_s = 0$ governs our system.

\subsection{Non-dimensionalization and perturbation expansion}

Scaling velocities and lengths over characteristic values $U$ and $a$ gives a characteristic surface pressure $\Pi_0=\eta_s^0U/a$, and the dimensionless surface momentum equation becomes:
\begin{equation}\label{nondim}
-\tilde{\bnabla}_s \tilde \Pi + \tilde{\bnabla}_s \bcdot \left[ \tilde \eta_s(\tilde \Pi) \left(\tilde{\bnabla}_s \bl{\tilde u}_s +\tilde{\bnabla}_s \bl{\tilde u}_s^T \right)\right]= \frac{1}{Bq}\pdv{\bl{\tilde v}}{\tilde z} \bigg|_{\text{z}=0}.
\end{equation}
In problems where a characteristic external force $F$ (instead of a velocity $U$) is specified, we can equivalently define $U=F/\eta_s^0$ or $\Pi_0=F/a$. The dimensionless surface viscosity is $\tilde \eta_s=\eta_s/\eta_s^0=e^{\Pi/\Pi_c}$ and the Boussinesq number
\begin{equation}
Bq=\frac{\eta_s}{\eta a}
\end{equation}
compares surface viscous stresses to subphase viscous stresses. Equation~\eqref{nondim} is essentially a 2D Stokes equation with non-constant viscosity forced by the traction from the bulk phase. Indeed, the solution of the corresponding Newtonian problem at large $Bq$ limits to 2D Stokes flow. Such a solution is classically known to be singular at large distances (due to the `Stokes paradox' \citep{Leal,Manikantan2020}). In the case of membranes or other viscous interfaces, traction from the bulk ultimately catches up to interfacial viscous stresses over the Saffman-Delbr{\"{u}}ck length \citep{Saffman1975,Manikantan2020} and regularizes the singularity. To keep the analysis tractable, we will work under the large-$Bq$ limit, and assume that the Saffman-Delbr{\"{u}}ck length is sufficiently large so that we can safely ignore the forcing term on the RHS of Eq.~\eqref{nondim}. We will appeal later to the physically realistic situation of a finite-size membrane to regularize the 2D Stokes description as described by Saffman \& Delbr{\"{u}}ck \citep{Saffman1975}.

In what follows, we drop the hats on dimensionless variables. The dimensionless governing equations are
\begin{equation}\label{nondim2}
\snabla\bcdot{\bl{\bsigma}} = -\snabla \Pi + \snabla\bcdot\left[\eta_s (\Pi)\bl{S}\right] = 0, \quad \bl{\snabla\cdot{u}} = 0,
\end{equation}
where $\bl{S}=\bnabla_s \bl{u}_s+\bnabla_s \bl{u}_s^T$ is twice the typical notation of the rate-of-strain tensor. Equation~\eqref{nondim2} is heavily nonlinear due to the dependence on $\eta(\Pi)$. To make analytical progress, we take a perturbative approach for large $\Pi_c$ or small departures of $\eta_s(\Pi)$ from $\eta^0_s$:
\begin{equation}
\eta_s(\Pi)=1 + \frac{\Pi}{\;\Pi_c} + \mathcal{O}(\Pi_c^{-2})= 1 + \beta \frac{\Pi}{\;\Pi_0}  + \mathcal{O}(\beta^2),
\end{equation}
where
\begin{equation}
	\beta = \frac{\Pi_0}{\Pi_c} = \frac{\eta_s^0 U}{\Pi_c a} = \frac{F}{\Pi_c a},
\end{equation}
is a dimensionless parameter that is small for weak pressure dependence of viscosity. Note that $\beta$ is positive when $\Pi$-thickening and negative when $\Pi$-thinning.

The momentum equation then expands to
\begin{equation}
\snabla\bcdot{\bl{\bsigma}}  = -	\snabla \Pi + \snabla \bcdot\bl{S} + \beta \snabla \bcdot\left[\Pi \bl{S}\right] + \mathcal{O}(\beta^2).
\end{equation}
In the weakly non-linear regime when $\beta \ll 1$, we asymptotically treat the interfacial velocity and surface pressure fields as regular expansions in $\beta$: 
\begin{subequations}
\begin{align}
	\bfu &= \bfu^{(0)} + \beta \bfu^{(1)} + \mathcal{O}(\beta^2), \\
	\Pi &= \Pi^{(0)} + \beta \Pi^{(1)} + \mathcal{O}(\beta^2),
\end{align}
\end{subequations}
so that the corresponding dimensionless stress tensor can be written as: 
\begin{subequations}\label{eq:stress}
\begin{align}
	\bsigma \hspace{10pt} & = \bsigma^{(0)}+\beta \bsigma ^{(1)} + \mathcal{O}(\beta^2), \\
	\bsigma^{(0)} & = -\Pi ^{(0)} \bl{I + S}^{(0)}, \\
	\bsigma ^{(1)} & = -\Pi ^{(1)} \bl{I + S}^{(1)} + \Pi ^{(0)} \bl{S}^{(0)} .
\end{align}
\end{subequations}
We note that the non-Newtonian term in the $\mathcal{O}(\beta)$ stress tensor $\bsigma^{(1)}$ depends on the leading-order solution. The momentum and mass conservation equations become: 
\begin{subequations}
\begin{align}
	\snabla \bcdot{\bsigma^{(0)}} &= \snabla \bcdot \bfu^{(0)}=0 \\
	\snabla \bcdot {\bsigma^{(1)}} &=\snabla \bcdot \bfu^{(1)} =0	\label{eq:orderbeta}
\end{align} 
\end{subequations}
We will first solve the leading-order Newtonian problem for a disk placed in arbitrary background surface flow. Then, we use the resulting stress field as the heterogeneity in the governing equation in $\mathcal{O}(\beta)$ to obtain the correction at $\mathcal{O}(\beta)$.

\subsection{Isolated disk driven in a background flow}
In building towards a description of the coupled dynamics of multiple disks, we first examine the Newtonian response of a single disk driven by an external force while placed in a background flow. We hope to then port the intuition as well as mathematical results to the case of a dilute 2D suspension of interacting disks by summing over the pair-wise hydrodynamic interactions. Such a Newtonian solution describes the leading order flow corresponding to ${\bsigma^{(0)}}$, $ \bfu^{(0)}$, and $\Pi^{(0)}$.

We analyze the translation of an isolated particle driven by a force $\bl{F}_p$ in the non-Newtonian interface with an imposed linear velocity field given by: 
\begin{equation}
	\bl{V(x) = V}_0 + \bl{x} \bcdot \boldsymbol{\Gamma}, 
\end{equation}
where $\boldsymbol{\Gamma}= \snabla \bl{V}$ is the imposed background velocity gradient. Let $q(x)$ and $\bl{A}=\snabla \bl{V} + (\snabla \bl{V})^T$ be the surface pressure and surface rate-of-strain fields associated with the imposed background surface flow. Meanwhile, we set $\Pi$ and $\bl{S}$ to be the pressure and rate of strain due to forced translation of the disk due to an applied external force $\bl{F}_p$. 

The stress fields at the leading and perturbed order, following Eq.~\eqref{eq:stress}, then take the forms
\begin{align}
	&\bsigma^{(0)}  = -(\Pi ^{(0)} + q) \bl{I}_s + (\bl{S}^{(0)} + \bl{A} ), \\
	&\bsigma ^{(1)}  = -\Pi ^{(1)}\bl{I}_s + \bl{S}^{(1)} +(\Pi ^{(0)} + q)(\bl{S}^{(0)} + \bl{A}).
\end{align}
It is advantageous in what follows to have disturbance fields decay to zero far from the particle. So, we define disturbance flow variables $\bl{\hat u }$, $\hat \Pi$, and $\hat \bsigma$ by subtracting imposed background flow variables from the total flow variables: 
\begin{subequations}
\begin{align}
	&\bl{\hat u(x) = u(x) - v(x) },\\ 
	&\hat \Pi (\bl{x}) 	= \Pi_t - q(\bl{x}), \\
	&\hat \bsigma (\bl{x}) = \bsigma (\bl{x}) - \btau(\bl{x}),
\end{align}
\end{subequations}
where $ \btau=-q \bl{I}_s + \bl{A} +\beta q \bl{A}$. These disturbance variables will also satisfy the momentum conservation and continuity equations and can be asymptotically expanded as a regular expansion in $\beta$: 
\begin{subequations}
\begin{align}
	\bl{\hat u} &= \bl{\hat u}^{(0)} + \beta  \bl{\hat u}^{(1)} +\ldots \\
	\hat \Pi &= \hat\Pi^{(0)} + \beta \hat\Pi^{(1)} +\ldots\\
	\hat \bsigma &= \hat\bsigma^{(0)} + \beta \hat\bsigma^{(1)} +\ldots
\end{align}
\end{subequations}

The $\mathcal{O} (1) $ disturbance problem corresponds to the linear Newtonian solution and can be written as the sum of disturbance velocity due to a translating disk and disturbance velocity due to the presence of a disk in an ambient flow. Using standard singularity methods of 2D Stokes flow \citep{Pozrikidis,Kim}, this velocity field can be shown to be 
\begin{equation}\label{leadingvel}
\begin{split}
	\bl{\hat u}^{(0)}(\bl{x}) = \left[2\left(-\ln{(r)}\bl{I} + \frac{\bl{xx}}{r^2}\right)+\left(\frac{\bl{I}}{r^2}-\frac{2\bl{xx}}{r^4}\right) \right]\bcdot \frac{\bl{F}_p}{4\pi} \\
- \left[\frac{\bl{xxx}}{r^4} + \frac{1}{2}\left(\frac{\bl{I x}}{r^4} - \frac{2\bl{xxx}}{r^6}\right)\right] \boldsymbol{:}\bl{A} .
\end{split}
\end{equation}
The terms proportional to $\bl{F}_p$ represent flow due to forced translation, whereas those proportional to the imposed velocity gradient via $\bl{A}$ are disturbances to background flow due to the presence of the disk. Note that the leading term due to the external force, corresponding to the 2D stokeslet, has a logarithmic singularity that is tied to the Stokes paradox mentioned previously. In reality, flow in the plane of the membrane is governed by 3D hydrodynamics ($ \bl{\hat u}^{(0)} \propto 1/r $) beyond the Saffman-Delbr{\"{u}}ck length, decaying to zero as the viscous traction from the subphase becomes dominant in the momentum balance. We will safely assume that the membrane is finite and within the momentum crossover or Saffman-Delbr{\"{u}}ck length, and this justifies the use of Eq.~\eqref{leadingvel} in capturing  membrane hydrodynamics when $Bq \gg 1$. 

The leading-order pressure disturbances also enter the $\mathcal{O}(\beta)$ governing equation. Like with the velocity, the Newtonian response can be written as a sum of contributions due to external force and imposed background velocity:
\begin{equation}\label{leadingpressure}
\hat \Pi ^{(0)}(\bl{x}) = \left[\frac{4\bl{x}}{r^2}\right]\bcdot \frac{\bl{F}_p}{4\pi} -\left[\frac{2\bl{xx}}{r^4}\right] \boldsymbol{:} \bl{A}.
\end{equation}
Force and torque balance ($\bl{F}_\text{net}=\bl{F}_p$ and $\bl{T}_\text{net}=\bl{0}$) with these Newtonian fields gives the translational velocity $\bl{U}_p=\bl{V}_0+\bl{F}_p/4\pi$ and rotational velocity $\boldsymbol{\Omega}_p = \snabla \times \bl{V}$ at leading order.

\subsection{Reciprocal theorem: non-Newtonian problem}
With the leading-order solution now known, we can write the non-Newtonian stress tensor at $\mathcal{O}(\beta)$ as
\begin{equation}
	\hat \bsigma ^{(1)}  = -\Pi ^{(1)} \bl{I} + \bl{S}^{(1)} +(\Pi ^{(0)} + q)(\bl{S}^{(0)} + \bl{A}) - (q\bl{A}),
\end{equation}
The effect of pressure-dependent surface viscosity could yield a non-zero $\ford{\bl{\hat F}}$ at $\mathcal{O}(\beta)$, given by: 
\begin{equation}
	\bl{\ford{\hat F}} = \int \hat{\bl{ n}} \bcdot \hat \bsigma ^{(1)} dl
\end{equation} 
where, $\bl{\hat n}$ is the normal vector pointing from the disk into the 2D fluid and $l$ denotes the boundary of the domain. As we have set up velocity field to decay far from the disk, this boundary is simply the perimeter of the disk. Without loss of generality, we choose to constrain the particle to its leading-order motion and determine the force or torque generated at $\mathcal{O}(\beta)$, i.e., $\bl{U}_p^{(1)}=\boldsymbol{\Omega}_p^{(1)}=\bl{0}$. The governing equations at this order, Eq.~\eqref{eq:orderbeta}, thus satisfy the boundary conditions
\begin{align}
	\bl{\hat u} ^{(1)}& = \bl{0} \quad \text{for} \quad \abs{\bl{x}} \leq 1,\\
	\bl{\hat u} ^{(1)}& \rightarrow  \bl{0} \quad \text{as} \quad \abs{\bl{x}}  \rightarrow \infty	.	
\end{align}

Evaluating the non-linear effects on the disk involves solving the $\mathcal{O}(\beta)$ problem for $\bl{\hat u} ^{(1)}$. This is an inhomogeneous Stokes equation that is extremely difficult to solve analytically. Instead, we make use of the Lorentz reciprocal theorem \cite{Masoud2019} to get around solving for the full flow and pressure fields using standard methods. This framework is well-established, and has previously been employed to evaluate perturbative solutions to several classes of weakly non-Newtonian problems involving viscoelasticity \cite{Vishnampet2012,HMPRSA,Chen2021}.

To set up the reciprocal theorem, we first need an auxiliary solution $\{\aux{\Pi},\;\bl{\aux{u}}\;\aux{\bsigma}\}$ that satisfies the homogeneous Stokes problem. We choose the solution to the translation of a disk in a 2D Newtonian interface with a velocity $\bl{\aux{{U}}}$. This solution is standard and is readily given by the singularity solutions corresponding to steady translation (i.e., the parts containing $\bl{F}_p$) in Eqs.~\eqref{leadingvel} and \eqref{leadingpressure}, with $\bl{\aux{{U}}}$ replacing $\bl{F}_p/4\pi$ in both cases. This Newtonian problem satisfies $\snabla \bcdot{\aux{\bsigma}}= \bl{0}$.

The reciprocal relation can be derived starting with the following vector relations:
\begin{subequations}
\begin{align}\label{eq:recip1}
\snabla \bcdot \left( {\aux{\bsigma}} \bcdot \bl{\hat u} ^{(1)} \right) = {\aux{\bsigma}} \boldsymbol{:} \bnabla\bl{\hat u} ^{(1)}, \\
\snabla \bcdot \left( \hat\bsigma^{(1)} \bcdot \aux{\bl{ u}}  \right) =\hat \bsigma^{(1)} \boldsymbol{:} \bnabla\aux{\bl{ u} }.\label{eq:recip2}
\end{align}
\end{subequations}
Subtracting Eq.~\eqref{eq:recip2} from Eq.~\eqref{eq:recip1} and integrating over the entire 2D fluid domain gives:
\begin{equation}\label{eq:recip3}
\begin{split}
\int \snabla \bcdot \left( {\aux{\bsigma}} \bcdot \bl{\hat u} ^{(1)} \right)  - \snabla \bcdot \left( \hat\bsigma^{(1)} \bcdot \aux{\bl{ u}}  \right) dS \\= \int {\aux{\bsigma}} \boldsymbol{:} \bnabla\bl{\hat u} ^{(1)} - \hat \bsigma^{(1)} \boldsymbol{:} \bnabla\aux{\bl{ u} } ~dS.
\end{split}
\end{equation}
The LHS of Eq.~\eqref{eq:recip3} can be simplified using the divergence theorem and force balance to give
\begin{equation}\label{eq:recip4}
\begin{split}
-\int \hat{\bl{n}} \bcdot \aux{\bsigma} \bcdot \bl{\hat u} ^{(1)} ~dl  + \int \hat{\bl{n}} \bcdot \hat\bsigma^{(1)} \bcdot \aux{\bl{ u}}  ~dl \\= -\aux{\bl{F}} \bcdot \bl{\ford{U}} +\bl{\ford{F} \bcdot \aux{U}} .
\end{split}
\end{equation}
Since we constrain the particle to its leading-order motion, the particle velocity at $\mathcal{O}(\beta)$ vanishes: $\bl{\ford{U}}=\bl{0}$. Then, using Eq.~\eqref{eq:recip4} for the LHS in Eq.~\eqref{eq:recip3} and simplifying gives
\begin{equation}
	\label{eq:recipfinal}
	\begin{split}
	\bl{\ford{F} \bcdot \aux{U}} = -\int \left[(\Pi ^{(0)} + q)(\bl{S}^{(0)} + \bl{A}) - q\bl{A}\right] \boldsymbol{:}\snabla \bl{\aux{u}} ~dS.
	\end{split}
\end{equation}
Importantly, Eq.~\eqref{eq:recipfinal} uses only the zeroth order (Newtonian) solution to solve a first order (non-Newtonian) problem.

As we are interested in far-field effects in dilute suspensions, we seek to incorporate leading order effects of the the disturbance velocity due to other disks as the imposed background flow. Over the lengthscale of a particle, this disturbance is primarily a unidirectional shear flow (so that $q(\bl{x})=0$). With this assumption, the integration in  Eq.~\eqref{eq:recipfinal} is still cumbersome but can be be performed analytically. Additionally, integrals of the leading order flow field involve the logarithm that diverges at large distances as noted before. We therefore restrict our fluid domain to a (dimensionless) lateral membrane size $R$ that is smaller that or comparable to the Saffman-Delbr{\"{u}}ck length: this corresponds to the finite membrane size resolution of the Stokes paradox \citet{Saffman1975}. With the velocity and pressure fields of the auxiliary Newtonian problem fully known, we finally get:
\begin{subequations}\label{eq:recipforce}
\begin{equation} \label{eq:recipforce1}
	\bl{\ford{F}} = \zeta \bl{F}_p \bcdot \bl{A},
\end{equation}
\begin{equation} \label{eq:recipforce2}
	\zeta = 4 \left[ \ln\left(R\right) - \frac{1}{4} -\frac{3}{R^2}+\frac{11}{4R^4}-\frac{1}{R^6} \right],
\end{equation}
\end{subequations}
where all quantities are dimensionless. The logarithmic dependence on membrane size that is characteristic of a 2D calculation of a driven object (unless otherwise regularized by bulk fluid momentum at large distances) shows up here and dominates the nonlinear response. 

The force that arises due to surface-pressure-dependent surface viscosity becomes evident if we revert to dimensional variables. Putting everything together, the leading-order dimensional force on a driven particle in an ambient linear flow field  in this 2D fluid is
\begin{equation}\label{eq:netvel}
\bl{F}_{\text{tot}} = \bl{F}_p + \frac{\zeta \eta_s^0}{\Pi_c} \bl{F}_p \bcdot \bl{A}  + \mathcal{O}(\beta^2).
\end{equation}
This is the main result of this paper. As expected, the non-Newtonian correction vanishes as $\Pi_c\rightarrow\infty$. More useful in many-particle simulations or continuum descriptions (\S~\ref{sec:collective}) is the net translational velocity resulting from such a force. Using the leading-order translational mobility $\bl{M}=\bl{I}_s/4\pi\eta_s^0 = M_0 \bl{I}_s$, and acknowledging the uniform component $\mathbf{V}_0$ of the background velocity, we find
\begin{equation}\label{eq:netvel2}
	\mathbf{U}_p = \mathbf{V}_0 + M_0 \left[ \mathbf{I}_s + \frac{\zeta \eta_s^0}{\Pi_c} \mathbf{A}   \right]\bcdot\mathbf{F}_p + \mathcal{O}(\beta^2).
\end{equation}
Notably, the correction proportional to $\mathbf{A} \bcdot \mathbf{F}_p$ is not necessarily along the direction of the driving force $\mathbf{F}_p$, thus setting the stage for lateral motion and concentration fluctuations in large-scale `suspensions' of trans-membrane inclusions.

\subsection{Physical interpretation and direction of motion}

\begin{figure}[t]
		\includegraphics[width=\linewidth]{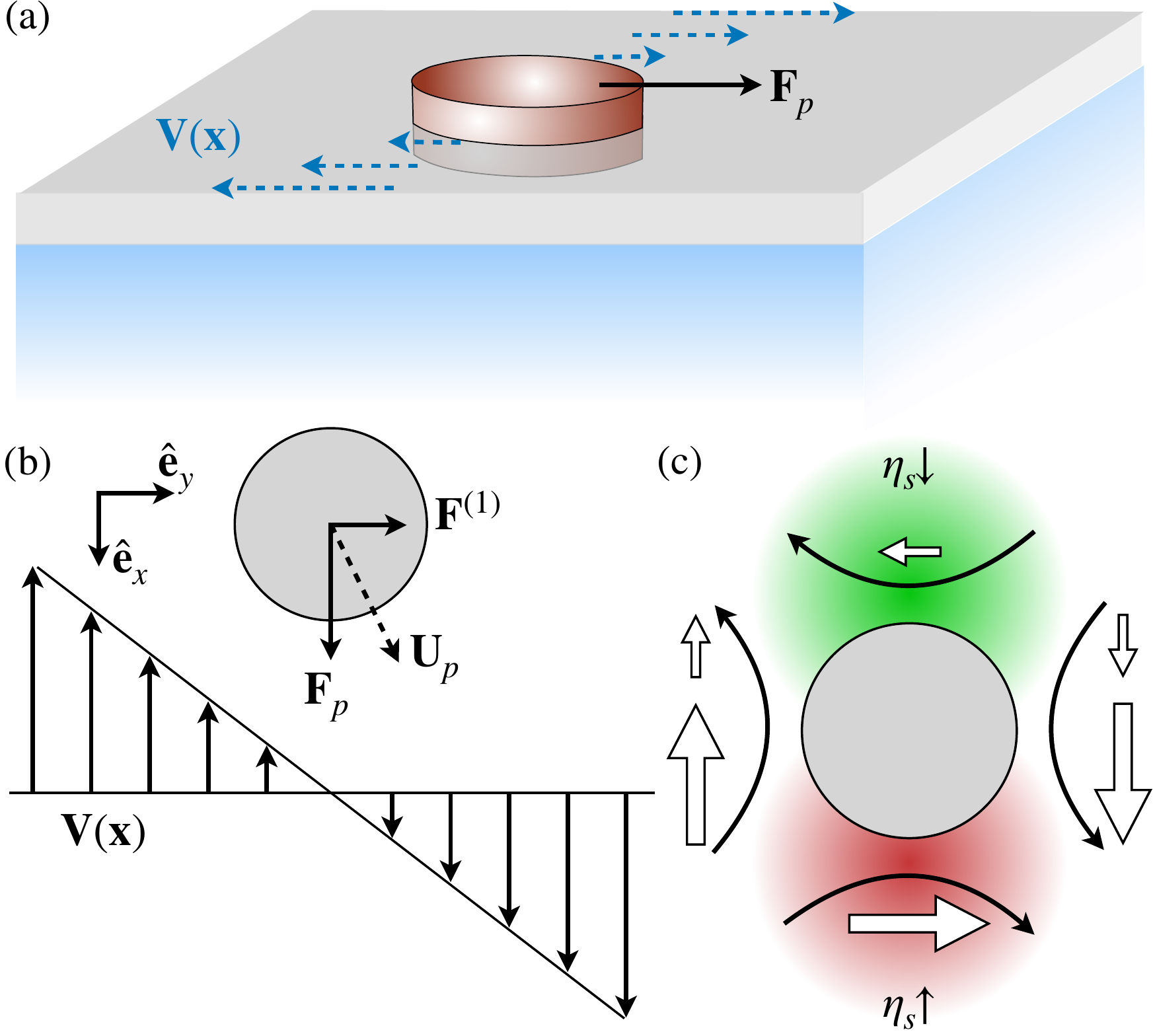}
		\caption{Physical mechanism leading to transverse migration in a driven particle placed in shear flow within the membrane as shown in (a). Background shear  as shown in (b) sets up an extensional flow as shown (c). The external force ${\bf F}_p$ on the particle also sets up pressure fields that lead to increase (in red) in viscosity ahead of the particle and decrease (in green) behind the particle. The resulting in-plane viscous stresses are proportional to the local viscosity and are asymmetric as indicated with the solid arrows. This results in a net force  ${\bf F}^{(1)}$ to the right in this particular configuration. }
		\label{fig:shear_mech}
	\end{figure}
	
We will first aim to mechanistically understand the direction of motion due to the non-Newtonian effect. A simple case is a background shear flow that is along the direction of the driving force (Fig.~\ref{fig:shear_mech}). This geometry is particularly relevant as it depicts the dominant effect of disturbance fields from neighboring particles, and is a common setup in classic studies of 3D sedimenting suspension mechanics \cite{Vishnampet2012,Koch1989}. In our case, we will take $\hat{\bl{e}}_x$ to be the direction along which  a disk embedded within a membrane is driven with an external horizontal force, so that $\bl {F}_p=F_p \hat{\bl{e}}_x$. Simultaneously, the 2D fluid is subject to an in-plane shear flow $\bl{v(x)} = \dot\gamma y \,\hat{\bl{e}}_x$ centered at the particle, so that the background strain tensor is
\begin{equation}
\bl{A} = \snabla\bl{v}+\snabla \bl{v}^T =   \dot\gamma (\hat{\bl{e}}_x \hat{\bl{e}}_y +\hat{\bl{e}}_y\hat{\bl{e}}_x).
\end{equation}
Then, the resulting net velocity following Eq.~\eqref{eq:netvel2} is
\begin{equation}
	\mathbf{U}_p =  M_0 F_p \hat{\bl{e}}_x +   \frac{\zeta\eta_s^0 \dot \gamma M_0 F_p}{\Pi_c} \, \hat{\bl{e}}_y.
\end{equation}
For $\Pi$-thickening suspensions, $\Pi_c>0$, and so the disk drifts in the $y$ direction in addition to being driven as expected in the $x$ direction by the applied force.

The direction of the non-Newtonian force can be gleaned by examining the nature of the pressure perturbation and disturbance flow around the disk. A shear flow can be locally decomposed into an extension and a rotation. The extension associated with the supplied shear is shown in Fig.~\ref{fig:shear_mech}(c). In a Newtonian medium, such an extensional flow does not generate a net force on the particle as surface viscous stresses around the perimeter of the particle cancel out. In a pressure-thickening fluid, however, the dipolar surface pressure field generates an associated surface viscosity field $\eta_s(\Pi)$ that makes the 2D medium more viscous ahead of the disk in the direction of $\bl {F}_p$ and less viscous behind it. The resulting surface viscous stresses are asymmetric around the particle, leading to a net force due to the extensional disturbance flow. The local surface viscous stresses are qualitatively illustrated in Fig.~\ref{fig:shear_mech}(c), which suggests that the resulting net force is to the right in this configuration.

This intuitive understanding of the interplay between extension due to background flow and local viscosity variations helps build up to more sophisticated models for coordinated behavior of collections of proteins or particles embedded within a membrane. In the following section, we take the mathematical result for the corrective force $\bl{\ford{F}}$ and the mechanistic picture drawn here to quantify such collective dynamics using mean-field kinetic models and particle simulations.

\section{Collective dynamics: Concentration Instability and Crystallization}\label{sec:collective}

\subsection{Mean-field model for dilute suspensions}

\begin{figure}[]
	\includegraphics[width=\linewidth]{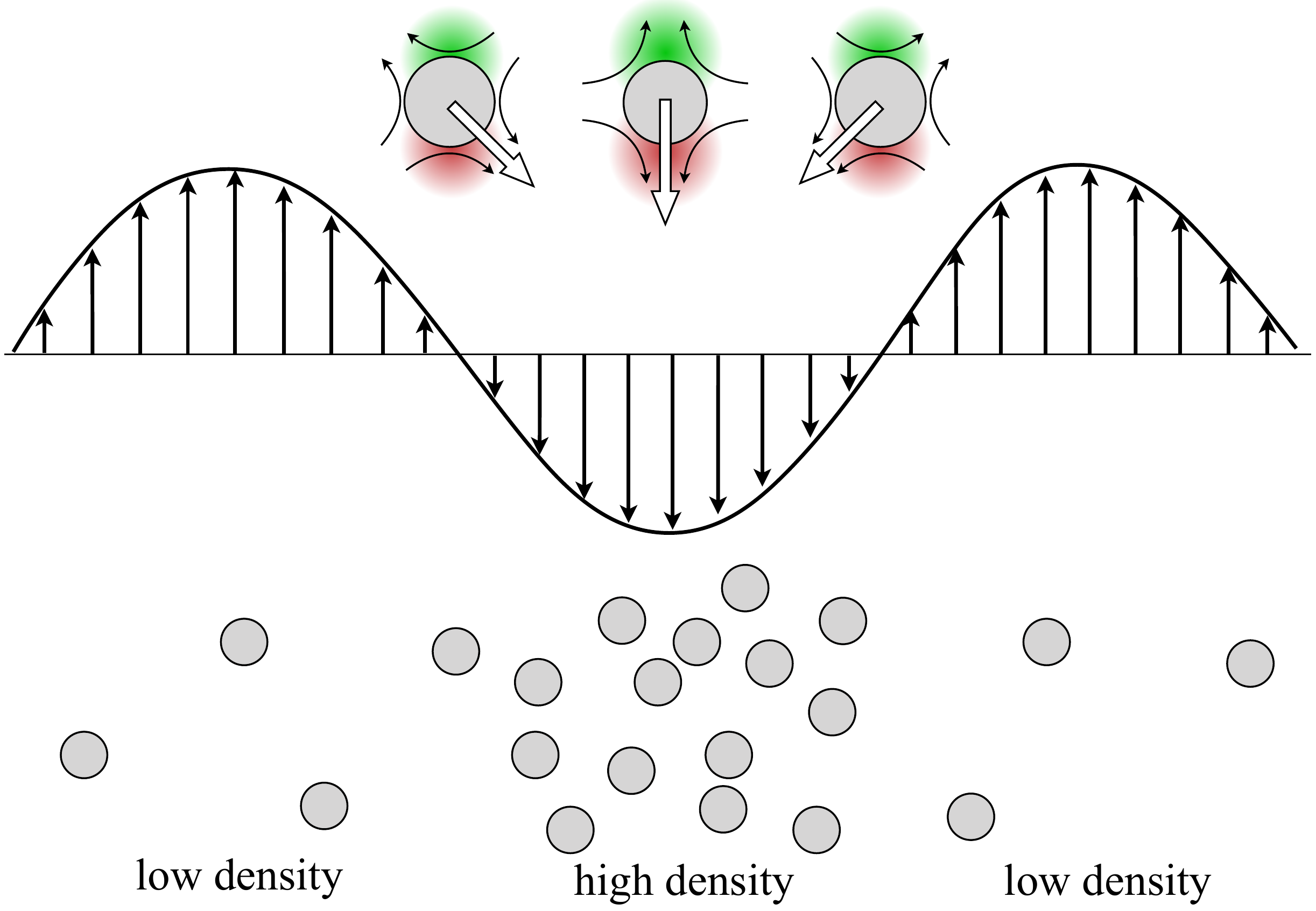}
	\caption{Mechanism for suspension instability. A local increase in particle concentration drives a shear flow that interacts with the non-uniform viscosity field around each particle in a manner that draws particles towards regions of high concentration. This magnifies the concentration fluctuation and the instability grows.}
	\label{fig:susp_mech}
\end{figure}

While the model for a single particle in a weakly non-Newtonian membrane is analytically tractable, the collective behavior of large-scale `2D suspensions' is much more complicated. Nevertheless, the direction of the corrective force derived above suggests a mechanism for large-scale aggregation of particles driven in a membrane. Simply put, the disturbance flow associated with the motion of each inclusion drives a local shear flow around every other inclusion. Cross-streamline motion within such a flow must directly lead to effective hydrodynamic drift towards regions of higher concentration of particles (Fig.~\ref{fig:susp_mech}). We intend to describe the concentration instability that results from such interactions.

We will first turn to a continuum description to glean physical features and stabilizing mechanisms of such an instability. We will follow the classic approach of \citet{Koch1989} in studying 3D suspensions of sedimenting rods, which has since been widely adapted to examine viscoelastic fluids \cite{Vishnampet2012}, active suspensions \cite{Saintillan2008}, and flexible fibers \cite{Manikantan2014}. 

We begin with a continuous variable $c(\bfx,t)$ that describes the concentration of particles embedded within this 2D medium at position $\bfx$ at time $t$. We will use $n$ to describe the mean number density (per unit area) on a interface of area $A$ such that
\begin{equation}
\frac{1}{A}\int c(\bfx,t)\,dA =n.
\end{equation}
Then, the evolution of local concentration is described by a conservation equation
\begin{equation}\label{eq:MFcons}
\frac{\partial c}{\partial t} + \snabla \bcdot (\dot{\bfx}\, c ) = D \nabla_s^2 c,
\end{equation}
where $\dot{\bfx}$ is the flux velocity and $D$ is a diffusivity that may represent Brownian fluctuations or hydrodynamic diffusion. We assume that the diffusivity is constant and isotropic, although such an assumption can be relaxed without qualitative changes to the conclusions drawn below. The flux velocity can be written as a combination of the background velocity field and the response of the particle to the applied force following Eq.~\eqref{eq:netvel2}:
\begin{equation}\label{eq:MFflux}
\dot{\bfx} = \bfu(\bfx) + M_0 \left[\bl{I}_s+ \frac{\zeta \eta_s^0}{\Pi_c} \left( \snabla\bfu + \snabla\bfu^T \right)  \right]\bcdot \bl{F}_p.
\end{equation}
Written this way, the conservation equation \eqref{eq:MFcons} captures the Newtonian response of each particle to the external force (via the term $M_0\bl{F}$, the non-Newtonian correction due to weak pressure-dependence (via the term proportional to $\zeta$), and the disturbance field due to the presence of neighboring particles at the location of each particle (via the background velocity field $\bfu(\bfx)$). This disturbed fluid velocity field is still unknown, and obeys
\begin{equation}\label{stokesinhomog}
-\snabla \Pi + \eta_s^0 \nabla_s^2 \bfu + \frac{\eta_s^0}{\Pi_c} \snabla \bcdot (\Pi  \bl{S}) + \bl{F}_p [c(\bfx)-n] = \bl{0},
\end{equation}
which is the 2D momentum equation perturbed by the external force at the location of the particles.

We will follow past works \cite{Vishnampet2012} to first argue that the nonlinear term in Eq.~\eqref{stokesinhomog} is sub-dominant in determining the flux velocity if the suspension is dilute and concentration fluctuations are small. We will stay in the regime of small perturbations in local concentration away from the uniform value $n$, so that $c(\bfx,t)=n+\epsilon c'(\bfx,t)$, with $|\epsilon| \ll  1$ and $ c'(\bfx,t)=\mathcal{O}(n)$. We will also expand $\bfu$ as a perturbation in $\beta$, so $\bfu=\bfu^{(0)}+\beta \bfu^{(1)}+\mathcal{O}(\beta^2)$ like in  previous sections. The leading-order disturbance velocity $\bfu^{(0)}$ then has characteristic magnitude $U_0\sim \epsilon F_p n L^2/\eta_s^0$ where $L$ is the suspension length scale (typically, the extent of the membrane or interface). The perturbative correction $\bfu^{(1)}$, by design, has characteristic scale $U_1 \sim \beta U_0$. We can compare this disturbance velocity field correction to the direct contribution due to the non-Newtonian effect obtained in Eq.~\eqref{eq:netvel2} which scales like $U_c \sim \beta F_p/\eta_s^0$. So, the nonlinear term in Eq.~\eqref{stokesinhomog} is relevant only if $U_1$ is comparable to $U_c$. However, $U_1/U_c\sim U_0\eta_S^0/F_p \sim \epsilon nL^2$. In a dilute 2D suspension, the area fraction of particles is small ($nL^2\ll1$) and the associated perturbation to linear stability is also small ($\epsilon\ll 1$). The non-Newtonian correction in equation \eqref{stokesinhomog} can therefore be safely neglected, and the disturbance field due to concentration fluctuations follow 
\begin{equation}\label{stokesinhomog_simp}
\snabla \Pi = \eta_s^0 \nabla_s^2 \bfu + \bl{F}_p [c(\bfx)-n],
\end{equation}
accurate to $\mathcal{O}(\epsilon \beta)$.

We wish to examine the linear stability of such a system to perturbations in concentrations away from the uniform state of $c(\bfx)=n$. The corresponding base state fluid velocity and surface pressure fields are $\bfu=\bl{0}$ are $\Pi(\bfx)=\Pi_0$, respectively. Perturbing the concentration as $c=n+\epsilon c'(\bfx,t)$ where $c'=\mathcal{O}(n)$ generates associated fluid fields $\bfu=\epsilon\bfu'(\bfx,t)$ and $\Pi=\Pi_0+\epsilon \Pi'(\bfx,t)$. Using these perturbed fields in the conservation equation \eqref{eq:MFcons} and linearizing gives the following form at $\mathcal{O}(\epsilon)$:
\begin{equation}\label{cons_linear}
\frac{\partial c'}{\partial t} + \frac{nM_0\zeta \eta_s^0}{\Pi_c} \nabla_s^2 \bfu' \bcdot \bl{F}_p+ M_0 \bl{F}_p \bcdot \snabla c' = D \nabla_s^2 c'.
\end{equation}
For stability analysis, we will consider normal modes of the form $c'=\tilde{c}(\bfk)\exp[i\bfk\bcdot\bfu+\sigma t ]$ and $\bfu'=\tilde{\bfu}(\bfk)\exp[i\bfk\bcdot\bfu+\sigma t ]$ with a 2D wavevector $\bfk$. The forced Stokes equation can be solved by applying Fourier transforms and using standard methods of projecting perpendicular to the pressure term \citep{Vishnampet2012,Hasimoto1959,Manikantan2020} to eliminate $\tilde{\bfu}(\bfk)$ in favor of $\tilde{c}(\bfk)$:
\begin{equation}
\tilde{\bfu}(\bfk)= \frac{1}{\eta_s^0 k^2} \left(\bl{I}_s-\hat{\bfk}\hat{\bfk} \right) \bcdot \bl{F}_p \tilde{c}(\bfk),
\end{equation}
where $k=|\bfk|$, and $\hat{\bfk}=\bfk/k^2$ is a unit wavevector. Using these Fourier coefficients in the linearized conservation equation \eqref{cons_linear} and simplifying gives
\begin{equation}
\begin{split}
\sigma \, \tilde{c}(\bfk) - \frac{nM_0\zeta }{\Pi_c} \bl{F}_p \bcdot \left(\bl{I}_s-\hat{\bfk}\hat{\bfk} \right) \bcdot \bl{F}_p \, \tilde{c}(\bfk) \\ + i k M_0 \bl{F}_p \bcdot \bfk \, \tilde{c} (\bfk) + k^2 D \, \tilde{c} (\bfk)= 0.
\end{split}
\end{equation}
Eliminating $\tilde{c}(\bfk)$ and defining $\theta=\cos^{-1}(\hat{\bl{e}}_x\bcdot \hat{\bfk})$ as the angle between the applied external force (taken to be along the $x$ direction, without loss of generality) and the wave vector, we obtain the real part of $\sigma$ representing the growth rate of concentration fluctuations:
\begin{equation}\label{dispersion_dim}
\sigma_R = \mathrm{Re}[\sigma] = \frac{nM_0\zeta F_p^2}{\Pi_c} \sin^2 \theta - k^2 D.
\end{equation}
Positive $\sigma_R$ indicates an exponentially growing perturbation based on the form of the normal modes. As $\zeta$, $\Pi_c$ and $M_0$ are all positive, the system is linearly unstable to concentration fluctuations  so long as $\theta$ is non-zero. 

\begin{figure}[]
	\includegraphics[width=\linewidth]{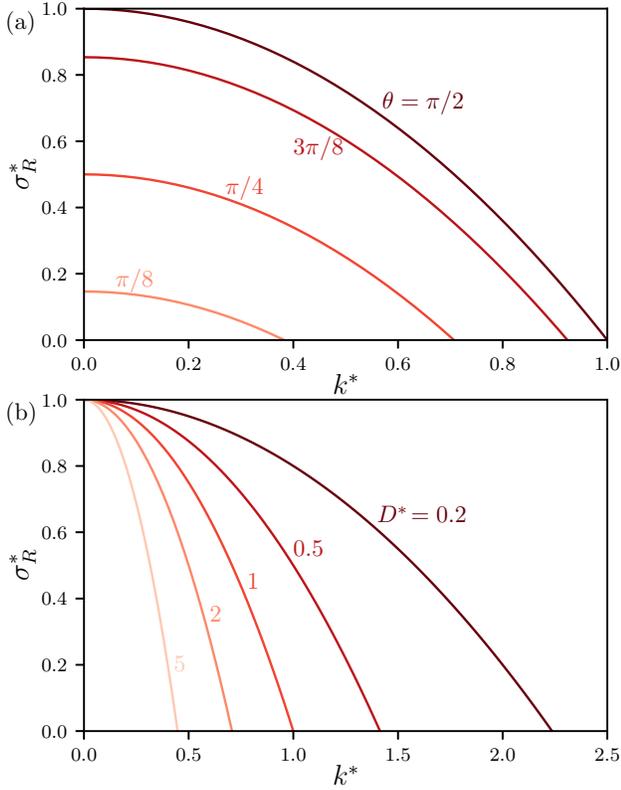}
	\caption{(a) Growth rate of concentration instability, nondimensionalized as $\sigma_R^*=\sigma \Pi_c/nM_0\zeta F_p^2)$, as a function of disturbance wavenumber $k^*=k/\sqrt{n}$. Curves represent different orientations of the wavevector relative to $\bl{F}_p \sim \hat{\bl{e}}_x$, with $\cos \theta=\hat{\bl{e}}_x\bcdot \hat{\bfk}$. (b) Diffusion acts to stabilize large wavenumber fluctuations, and a higher diffusivity suppresses a wider range of modes: here, $D^*=D \Pi_c/M_0\zeta F_p^2$.}
	\label{fig:disp}
\end{figure}

Figure~\ref{fig:disp} shows the dispersion relation from Eq.~\eqref{dispersion_dim} made dimensionless by scaling the growth rate over $nM_0\zeta F_p^2/\Pi_c$, the wave number over $\sqrt{n}$ and the diffusion constant over $M_0\zeta F_p^2/\Pi_c$. Such a driven suspension is evidently unstable to long-wavelength perturbations. Indeed, the growth rate is always largest at $k=0$, corresponding to concentration fluctuations that span the system size. Perturbations in the direction perpendicular to the driving force (so that $\theta=\pi/2$) are the most destabilizing, analogous to classic theories of sedimentation stability in 3D suspensions \citep{Koch1989,Saintillan2008,Manikantan2014}. The only stabilizing mechanism is diffusion, that suppresses large wave numbers (as shown in Fig.~\ref{fig:disp}b). As expected, the instability does not occur in the Newtonian limit as $\Pi_c \rightarrow \infty$.

The associated mechanism is illustrated in Fig.~\ref{fig:susp_mech}. A perturbation that increases local particle concentration drives a shear flow that interacts with the non-uniform viscosity field around each particle to generate a non-Newtonian correction to the drift velocity as shown in previous sections (see Fig.~\ref{fig:shear_mech}). In a suspension, this drift draws particles towards regions of high concentration. This magnifies the concentration fluctuation and the instability grows.

\subsection{Langevin description}

The mean-field model above describes short-term linear growth of concentration fluctuations and suggests a mechanism for aggregation. In what follow, we will demonstrate the long-term effect of the nonlinear force on collections of membrane inclusions using a Langevin description. We will consider $N$ disks embedded in a membrane with an infinite sub-phase, where each disk translates with a self velocity due to an applied force $\bl{F}_p$. This could be due to anchoring proteins on a membrane, or equivalently due to externally applied electrical, thermal, or optical forces on particles embedded within a complex interface. We assume that the suspension is sufficiently dilute such that pairwise interactions can be added to obtain net force on each disk. The trajectory of each disk is then determined by the Newtonian membrane hydrodynamic disturbance field due to the other $N-1$ disks, the correction due to non-Newtonian nature of the membrane as derived above, and Brownian fluctuations.

The corresponding Langevin equation for disk $i$ reads: 
\begin{equation} \label{eq:Langevin}
\frac{\partial \bl{x}_{i}}{\partial t}= M_0 F_p \hat{\bl{e}}_x + \bfu_i^{\text{Br}}  + \sum_{i\neq j}^{N} \left[\bfu^d_j(\bl{x}_i) + \bfu^{\text{St}}_{ij}  + M_0 \beta \bl{F}_{ij} \right].
\end{equation}
The first term on the RHS describes the translational response of disk $i$ to the applied force. We will take $\bl{F}_p=F_p \hat{\bl{e}}_x$ to be constant and along the $x$ direction. The second term describes Brownian fluctuations, which in general depend on the collective hydrodynamic motility of the system \cite{Oppenheimer2009}. Owing to the diluteness of our system, we only account for the local mobilities for each disk. This simplification decouples the fluctuation-dissipation relation of each disk from the rest so that the translational Brownian velocity satisfies:
\begin{eqnarray}\label{flucdiss}
	\avg{\bfu^{\text{Br}}(t)} = 0, \quad \avg{\bfu^{\text{Br}}(t) \bfu^{\text{Br}}(t')}= 2k_BT\bl{M}\delta (t-t'),
\end{eqnarray}
where $\bl{M}=M_0\bl{I}_s=\bl{I_s}/4\pi \eta_s^0$ is the Newtonian translational mobility of a disk. Using $U=F_p/\eta_s^0$ and $a/U$ as the characteristic scales for velocity and time, we obtain the dimensionless Brownian velocity from Eq.~\eqref{flucdiss} as
\begin{equation}
	\bfu_i^{\text{Br}} = \sqrt{\frac{\tilde{T}}{2\pi \Delta \tilde t}}~ \bl{w}, \qquad  \tilde{T}=\frac{k_B T}{F_p a},
\end{equation}  
where $\Delta \tilde{t}$ is the dimensionless time step and $\tilde{T}$ is a dimensionless temperature. The white noise vector $\bl{w}$ is populated by random numbers sampled from a normal distribution of mean 0 and variance 1 such that the fluctuation-dissipation relation Eq.~\eqref{flucdiss} is satisfied. 

The next term in the Langevin equation describes leading-order disturbance field at the location of disk $i$ due to an external force on neighboring disk $j$:
\begin{equation}
\bfu^d_j(\bl{x}_i) = \bl{G}_{ij} (\bfx_i-\bfx_j) \bcdot \bl{F}_p.
\end{equation}
Here, $\bl{G}_{ij}$ is the Green's function or 2D stokeslet for the Newtonian problem. This tensor is readily obtained from the term corresponding to $\bl{F}_p$ in Eq.~\eqref{leadingvel}. 

Although the suspension is dilute and particles are initially far from each other, fluctuations due to pair-wise addition of attractive hydrodynamic forces can bring disks within contact distance of each other. To prevent overlap, we add a soft repulsion between disks at short range. These repulsive excluded volume interactions are accounted via a steric velocity
\begin{equation}
	 \bfu^{\text{St}}_{ij}= U_s\frac{e^{-\alpha(r-a)}}{1+e^{-\alpha(r-a)}} \, \hat{\bl{r}}_{ij},
\end{equation}   
where $r=|\bfx_i-\bfx_j|$, and which acts along the unit vector $\hat{\bl{r}}_{ij}=(\bfx_i-\bfx_j)/r$ that connects disk $i$ to disk $j$. This soft repulsion decays exponentially over a length scale $\alpha^{-1}=\mathcal{O}(a)$ and $U_s$ is the contact velocity.

\begin{figure*}[t]
		\includegraphics[width=\linewidth]{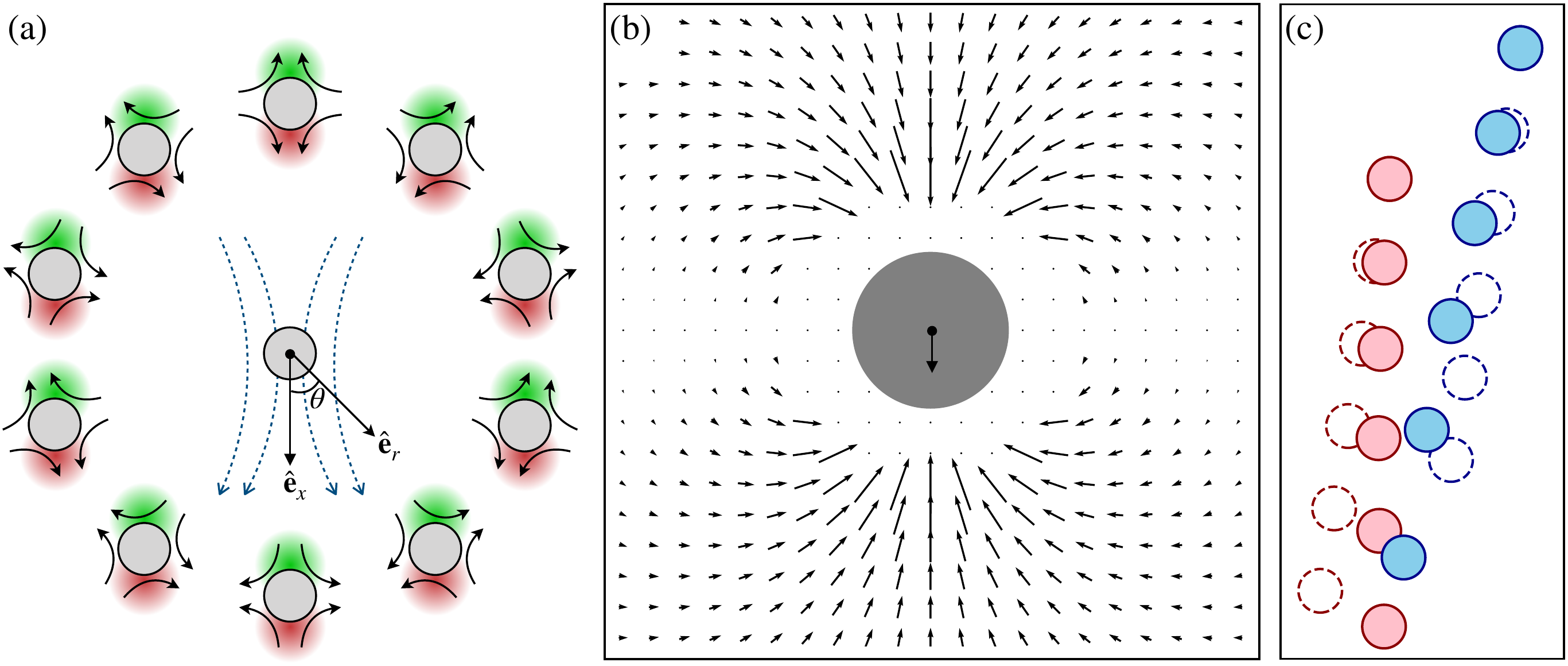}
		\caption{(a) Local extension field around particles due to the flow generated by a reference particle at the center. (b) Effective non-Newtonian interaction due to the interplay between local extension and non-uniform viscosity reveals a propensity to chain particles along the direction of the applied force. (c) Snapshots at equal intervals of time of a pair of particles translating in response to a downward force, simulated using the Langevin description. Dashed unfilled particles are the Newtonian reference case, which maintain relative separation and orientation. Solid filled circles account for pressure-dependent viscosity, which draws them together to chain up along the direction of the force.}
		\label{fig:chaining}
	\end{figure*}

The final term in the Langevin description is the non-Newtonian correction from Eq.~\eqref{eq:netvel2} obtained via the reciprocal calculation in section~\ref{sec:single}. This term also acts pair-wise as the local background velocity gradient felt by disk $i$ is that generated by the disturbance velocity associated with disk $j$. We time integrate the Langevin equation numerically using standard Brownian dynamics methods established for similar systems \citep{Manikantan2020PRL,Oppenheimer2019}.

\subsection{Effective hydrodynamic interactions and chaining}

Before discussing long-time features and hydrodynamic crystallization, we first take a closer look at the effective hydrodynamic interactions in large-scale suspensions to understand the mechanism of particle aggregation. The interplay between local fluid extension and non-uniform viscosity field around each particle dictates the effective hydrodynamic interactions that drive relative particle motion. Fig.~\ref{fig:chaining}(a) borrows from the classic work of \citet{Koch1989} to show local extension around a particle that is placed near a point force (corresponding to a reference particle) at the center. Each surrounding particle represents a possible location of a membrane anchor protein in the disturbance field induced by the reference particle. The orientation of the extensional field changes based on the location of the particle, but always acts to draw fluid towards the reference point force in the upper half and to draw fluid away from it in the lower half. 

Recalling that inclusions are also driven downward by external forces, the local viscosity field around each inclusion is simultaneously weakly perturbed. The particles experience a relatively higher viscosity ahead of them. The effective hydrodynamic force, or the resulting drift velocity, can then be determined by evaluating the asymmetric surface viscous stresses around each particle. Again, such extensional flows would not generate a net force on disks in a Newtonian system (akin to spheres in 3D fluids \cite{Koch1989}): the non-uniform viscosity around each inclusion breaks this symmetry.

We can readily evaluate the form of such an interaction. The flow field set up by the reference particle (the dashed lines in Fig.~\ref{fig:chaining}a) is given by the term proportional to the force in Eq.~\eqref{leadingvel}. For illustration, we consider only the leading-order disturbance field corresponding to a 2D stokeslet or a point force. The correction due to the interplay between the external forces and the strain rate corresponding to such a flow at a location $\bfx$ is given, following Eq.~\eqref{eq:recipforce}, by
\begin{equation}\label{eq:chain1}
\bl{\ford{F}} = \zeta \bl{F}_p \bcdot \bl{A} = \zeta \bl{F}_p \bcdot \left[ \frac{\bl{I}_s \bfx}{r^2} - \frac{2 \bfx\bfx\bfx}{r^4} \right] \bcdot \bl{F}_p,
\end{equation}
where $r=|\bfx|$. Taking $\bl{\ford{F}}= F_p\hat{\bl{e}}_x$ and denoting the radial unit vector originating at the reference particle as $\hat{\bl{e}}_r$, Eq.~\eqref{eq:chain1} becomes
\begin{equation}\label{eq:chain2}
\bl{\ford{F}} = \zeta F_p^2 \left( \frac{\cos \theta}{r} \hat{\bl{e}}_x - \frac{2 \cos^2 \theta}{r} \hat{\bl{e}}_r \right),
\end{equation}
where now $\cos\theta=\hat{\bl{e}}_x \bcdot \hat{\bl{e}}_r$.

This force leads to a velocity $M_0\bl{\ford{F}} $ at the location $\bfx$, which can be interpreted as an effective hydrodynamic interaction due to pressure-dependent viscosity. $M_0$ is the isotropic mobility of a disk, and the direction of the velocity follows that of $\bl{\ford{F}} $. This correction to the Newtonian-flow field around each particle, following Eq.~\eqref{eq:chain2}, is shown in Fig.~\ref{fig:chaining}(b). As expected, this correction vanishes in the horizontal line where $\theta=\pi/2$, where the extensional field due to a stokeslet is zero. The correction is non-zero at all other orientations and maintains a front-back symmetry relative to $\hat{\bl{e}}_x$. Notably, the direction of $\bl{\ford{F}} $ is such that it acts to draw neighboring particles into chains along the direction of external force.

\begin{figure*}[t]
		\includegraphics[width=\linewidth]{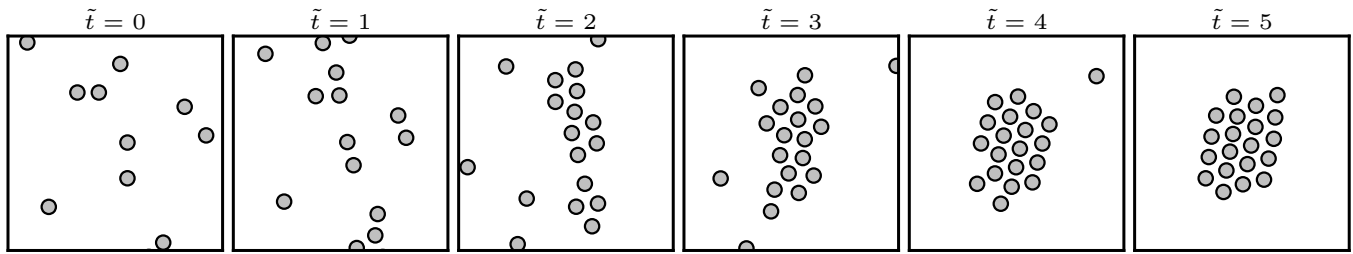}
		\caption{Time series of a 2D suspension of disks driven by an external (downward) force leading to hydrodynamic aggregation. Chains form initially which eventually give way to large scale hexatically ordered crystals. Here, $\beta=0.1$ and $\tilde{T}=0.001$.}
		\label{fig:sims}
	\end{figure*}

Such a mechanism of chaining is clearly seen in Fig.~\ref{fig:chaining}(c) in two-particle simulations following the Langevin description from the preceding section. While a pair of particles in a 2D Newtonian fluid would maintain relative separation and orientation as a they translate in response to an external force, the non-Newtonian correction derived here causes them to line up along the direction of the external force. We therefore expect particles to form chains. By analogy with molecular systems that display a long-ranged attraction and short-range repulsion, we expect clustering and aggregation when a large number of particles are present. 

Indeed, our simulations show that chains form initially but eventually give way to larger-scale stable clusters (Fig.~\ref{fig:sims}) with increasing $\beta$. Aggregates are stable and do not disintegrate for long simulation times (as observed till $\tilde t = 120$ in units of time made dimensionless over characteristic time $a\eta_s^0/F_p$, Fig.~ \ref{fig:sims}). We see crystalline order emerge as expected from analogous problems with `hydrosteric' interactions \citep{Oppenheimer2019}, i.e, long-ranged attraction and short-rainge repulsive interactions. Previous works have examined aggregation of driven transmembrane proteins due membrane curvature \citep{Reynwar2007,Dasgupta2017}, and inclusion size mismatch \cite{Bruinsma1996}. Our simulations show that stable hexatic clusters form as a result of the hydrodynamic attraction, physically representing a collection of membrane anchor points crystallizing for a potential biological advantage. 
	
\subsection{Aggregation and hexatic Order}

We will use particle pair proximity and resulting packing order to quantify the degree of aggregation in  this system. A straightforward metric is the $n$-th packing order parameter:
\begin{equation}
	\Psi^j_n=\frac{1}{n_j}\sum_{k} e^{in\theta_{kj}}.
\end{equation}   
$\Psi^j_n$ measures the orientation and packing order around particle $j$, where $n_j$ is the number of nearest neighbors and $\theta _{kj}$ is the angle between the line joining disks $k,j$ and the x-axis. The average local $\avg{\abs{\Psi_n}}$ order parameter quantifies the packing order, with 0 corresponding to an unordered aggregate and 1 representing a perfect $n$-th order aggregate. The local order parameter tells us whether each particle in an aggregate forms a $n$-th order lattice with its nearest neighbor. Note that the \emph{global} hexatic order parameter $|\avg{\Psi_n}|$ can indicate system-spanning crystalline structure\cite{Oppenheimer2019}. Our simulations are non-periodic and have a finite number of particles, and we confirm that no qualitative difference exists between the local and global order parameters for systems at the scales shown.

\begin{figure}[b]
		\includegraphics[width=\linewidth]{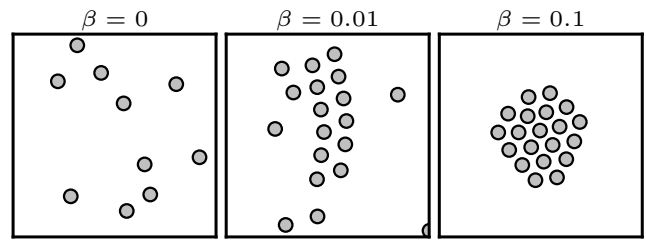}
		\caption{A stronger pressure-dependence of viscosity (smaller $\Pi_c$ or larger $\beta$) leads to faster aggregation and crystallization. Snapshots taken at $\tilde{t}=25$.}
		\label{fig:betas}
	\end{figure}

Given the 2D nature of our systems and the short-ranged repulsion, we always see hexatic ($n=6$) order formation for non-zero $\beta$. This is comparable to other works on 2D nonlinear hydrodynamics \cite{Oppenheimer2019,Grzybowski2000,Goto2015} that have shown that rigid inclusions rotating on a membrane form stable hexagonal crystals. The effective attractive potential in our case is orientation-dependent (Fig.~\ref{fig:chaining}b) and arises due to surface-pressure dependent rheology. As expected, disks self-assemble faster with increasing $\beta$ (Fig.~\ref{fig:betas}) corresponding to a stronger role of pressure-dependent surface viscosity.

\begin{figure}
		\includegraphics[width=\linewidth]{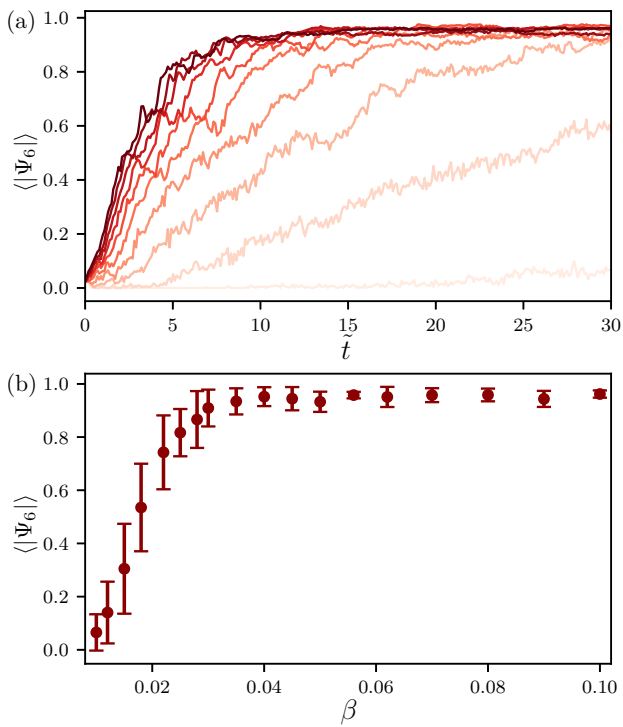}
		\caption{(a) Evolution of the hexatic order parameter with time. Darker hue indicates a larger $\beta$, and the curves correspond to $\beta=0.01$ to $\beta=0.1$ in steps of $0.01$. (b) Plateau value of $\langle | \Psi_6 | \rangle$ at $\tilde{t}=30$.}
		\label{fig:plateau}
	\end{figure}

We can also use the evolution  of  $\avg{\abs{\Psi_6}}$ to quantitatively describe the rate of assembly, and qualitatively describe the transition towards hexatic order. Hexagonal order of the disk crystals plateaus with time to attain a quasi-steady value (Fig.~ \ref{fig:plateau}a). The rate at which this plateau is reached, as well the magnitude of the plateau are clear indicators of departure from Newtonian behavior. The transition from disorder to crystalline order due to $\Pi$-dependent $\eta_s$ is evident in Fig.~\ref{fig:plateau}(b).

Finally, note that the strong thermal fluctuations can melt crystals and restabilize the suspension towards a uniform concentration. We have set the effective temperature $\tilde{T}=k_BT/F_p a =0.001$ in the results shown here, which corresponds to $F_p=\mathcal{O}(\text{pN})$ for particles or membrane inclusions in the size range of $a=\mathcal{O}(100\text{ nm})$. This is in the range of forces exerted on membranes by biological activity such as intracellular pinning and cytoskeleton-associated growth and locomotive forces \cite{Fogelson2014,Cohen2020}. Larger forces or bigger particles would only accentuate the effect, and we have tested that crystallization transition is qualitatively unchanged upon increasing thermal noise upto $\tilde{T}=0.01$. Increasing thermal fluctuations within this biologically relevant range slightly decreases the plateau value of $\avg{\abs{\Psi_6}}$: the rate at which the plateau value is reached for all $\beta$, however, remains unchanged.

\section{Conclusion}

Collective motility, aggregation, and crystallization of membrane proteins has distinct biological advantage in immune response, locomotion, and tubulation \cite{Bruinsma1996,Fogelson2014,Reynwar2007}. Previous works have examined aggregation that arises due to protein activity \cite{Oppenheimer2019,Manikantan2020PRL}, membrane curvature \citep{Reynwar2007,Dasgupta2017}, and inclusion size mismatch \cite{Bruinsma1996}. We have shown that a commonly measured class of non-Newtonian membrane viscosity can have the same effect, and that these non-trivial fluid dynamics can play critical roles in membrane organization principles \cite{Lingwood2010,Sezgin2017}. 

We have made several approximations to keep this complicated problem tractable. Neglecting the coupling to the bulk phase will have quantitative changes to the corrective force, especially for large membranes where 3D fluid stresses catch up over the Saffman-Delbr{\"{u}}ck crossover length. We expect this change to be minimal for finite membrane sizes or crowded system, as the bulk flow effect is screened at much shorter length scales, and so the qualitative picture remains unchanged. A more careful analysis of the coupling of a non-Newtonian membrane to a 3D fluid phase is left as future work. We have also chosen to develop a minimal point-particle-based Langevin approach for simulations. Our objective was to demonstrate the long-time dynamics with a minimal simulation model. We expect that high-fidelity computations will confirm these dynamics, and also go beyond the weakly nonlinear regime. We also recognize that not all inclusions within a membrane are driven: a real membrane hosts a collection of anchored, driven, active, and passive inclusions. Nor are inclusions disk-like. While shape does not significantly alter long-ranged membrane hydrodynamics, the near-field steric effects and packing order will shift to accommodate inclusion shape. Nonetheless, we expect that the physical intuition gained in this study will port over to future analyses of combinations of mobile and immobile inclusions

We end by noting that the mathematical framework and physical picture that emerges in this work is generalizable to a broader class of problems beyond lipids with pressure-dependent viscosity. The formulation using the reciprocal theorem is fairly general and can be extended to other common non-Newtonian constitutive relations. Force corrections due to shear-thinning \cite{Chen2021} and viscoelastic effects \cite{Vishnampet2012} follow their 3D analogs. The mean-field model can also be readily adapted to any form of the corrective force, opening a rich avenue of problems that can be viewed through the lens of this framework. Further, this model can be applied to a broader class of 3D fluids with pressure-dependent viscosity, such as those that appear in piezoviscous problems \cite{Hron2001}, high-pressure polymer melts \cite{Denn1981}, geophysical flows \cite{Schoof2007}, and crude oil mixtures \cite{Srinivasan2013}. 


\vspace{5mm}
\noindent {\bf Conflict of Interest:} The authors have no conflicts to disclose.

\vspace{5mm}
\noindent {\bf Acknowledgments:} We thank Ronald J. Phillips and Gregory H. Miller for insightful feedback on this work.

\vspace{5mm}


\begin{thebibliography}{60}%
\makeatletter
\providecommand \@ifxundefined [1]{%
 \@ifx{#1\undefined}
}%
\providecommand \@ifnum [1]{%
 \ifnum #1\expandafter \@firstoftwo
 \else \expandafter \@secondoftwo
 \fi
}%
\providecommand \@ifx [1]{%
 \ifx #1\expandafter \@firstoftwo
 \else \expandafter \@secondoftwo
 \fi
}%
\providecommand \natexlab [1]{#1}%
\providecommand \enquote  [1]{``#1''}%
\providecommand \bibnamefont  [1]{#1}%
\providecommand \bibfnamefont [1]{#1}%
\providecommand \citenamefont [1]{#1}%
\providecommand \href@noop [0]{\@secondoftwo}%
\providecommand \href [0]{\begingroup \@sanitize@url \@href}%
\providecommand \@href[1]{\@@startlink{#1}\@@href}%
\providecommand \@@href[1]{\endgroup#1\@@endlink}%
\providecommand \@sanitize@url [0]{\catcode `\\12\catcode `\$12\catcode
  `\&12\catcode `\#12\catcode `\^12\catcode `\_12\catcode `\%12\relax}%
\providecommand \@@startlink[1]{}%
\providecommand \@@endlink[0]{}%
\providecommand \url  [0]{\begingroup\@sanitize@url \@url }%
\providecommand \@url [1]{\endgroup\@href {#1}{\urlprefix }}%
\providecommand \urlprefix  [0]{URL }%
\providecommand \Eprint [0]{\href }%
\providecommand \doibase [0]{https://doi.org/}%
\providecommand \selectlanguage [0]{\@gobble}%
\providecommand \bibinfo  [0]{\@secondoftwo}%
\providecommand \bibfield  [0]{\@secondoftwo}%
\providecommand \translation [1]{[#1]}%
\providecommand \BibitemOpen [0]{}%
\providecommand \bibitemStop [0]{}%
\providecommand \bibitemNoStop [0]{.\EOS\space}%
\providecommand \EOS [0]{\spacefactor3000\relax}%
\providecommand \BibitemShut  [1]{\csname bibitem#1\endcsname}%
\let\auto@bib@innerbib\@empty
\bibitem [{\citenamefont {Phillips}\ \emph {et~al.}(2012)\citenamefont
  {Phillips}, \citenamefont {Kondev}, \citenamefont {Theriot},\ and\
  \citenamefont {Garcia}}]{rob}%
  \BibitemOpen
  \bibfield  {author} {\bibinfo {author} {\bibfnamefont {R.}~\bibnamefont
  {Phillips}}, \bibinfo {author} {\bibfnamefont {J.}~\bibnamefont {Kondev}},
  \bibinfo {author} {\bibfnamefont {J.}~\bibnamefont {Theriot}},\ and\ \bibinfo
  {author} {\bibfnamefont {H.}~\bibnamefont {Garcia}},\ }\href@noop {} {\emph
  {\bibinfo {title} {Physical biology of the cell}}}\ (\bibinfo  {publisher}
  {Garland Science},\ \bibinfo {year} {2012})\BibitemShut {NoStop}%
\bibitem [{\citenamefont {Van~Meer}, \citenamefont {Voelker},\ and\
  \citenamefont {Feigenson}(2008)}]{lip}%
  \BibitemOpen
  \bibfield  {author} {\bibinfo {author} {\bibfnamefont {G.}~\bibnamefont
  {Van~Meer}}, \bibinfo {author} {\bibfnamefont {D.~R.}\ \bibnamefont
  {Voelker}},\ and\ \bibinfo {author} {\bibfnamefont {G.~W.}\ \bibnamefont
  {Feigenson}},\ }\bibfield  {title} {\enquote {\bibinfo {title} {Membrane
  lipids: where they are and how they behave},}\ }\href@noop {} {\bibfield
  {journal} {\bibinfo  {journal} {Nature reviews Molecular cell biology}\
  }\textbf {\bibinfo {volume} {9}},\ \bibinfo {pages} {112--124} (\bibinfo
  {year} {2008})}\BibitemShut {NoStop}%
\bibitem [{\citenamefont {Saffman}\ and\ \citenamefont
  {Delbruck}(1975)}]{Saffman1975}%
  \BibitemOpen
  \bibfield  {author} {\bibinfo {author} {\bibfnamefont {P.~G.}\ \bibnamefont
  {Saffman}}\ and\ \bibinfo {author} {\bibfnamefont {M.}~\bibnamefont
  {Delbruck}},\ }\bibfield  {title} {\enquote {\bibinfo {title} {{Brownian
  motion in biological membranes.}}}\ }\href
  {https://doi.org/10.1073/pnas.72.8.3111} {\bibfield  {journal} {\bibinfo
  {journal} {Proceedings of the National Academy of Sciences}\ }\textbf
  {\bibinfo {volume} {72}},\ \bibinfo {pages} {3111--3113} (\bibinfo {year}
  {1975})}\BibitemShut {NoStop}%
\bibitem [{\citenamefont {Hughes}, \citenamefont {Pailthorpe},\ and\
  \citenamefont {White}(1981)}]{Hughes1981}%
  \BibitemOpen
  \bibfield  {author} {\bibinfo {author} {\bibfnamefont {B.~D.}\ \bibnamefont
  {Hughes}}, \bibinfo {author} {\bibfnamefont {B.~A.}\ \bibnamefont
  {Pailthorpe}},\ and\ \bibinfo {author} {\bibfnamefont {L.~R.}\ \bibnamefont
  {White}},\ }\bibfield  {title} {\enquote {\bibinfo {title} {{The
  translational and rotational drag on a cylinder moving in a membrane}},}\
  }\href {https://doi.org/10.1017/S0022112081000785} {\bibfield  {journal}
  {\bibinfo  {journal} {Journal of Fluid Mechanics}\ }\textbf {\bibinfo
  {volume} {110}},\ \bibinfo {pages} {349} (\bibinfo {year}
  {1981})}\BibitemShut {NoStop}%
\bibitem [{\citenamefont {Choi}\ \emph {et~al.}(2011)\citenamefont {Choi},
  \citenamefont {Steltenkamp}, \citenamefont {Zasadzinski},\ and\ \citenamefont
  {Squires}}]{Choi2011}%
  \BibitemOpen
  \bibfield  {author} {\bibinfo {author} {\bibfnamefont {S.~Q.}\ \bibnamefont
  {Choi}}, \bibinfo {author} {\bibfnamefont {S.}~\bibnamefont {Steltenkamp}},
  \bibinfo {author} {\bibfnamefont {J.~A.}\ \bibnamefont {Zasadzinski}},\ and\
  \bibinfo {author} {\bibfnamefont {T.~M.}\ \bibnamefont {Squires}},\
  }\bibfield  {title} {\enquote {\bibinfo {title} {{Active microrheology and
  simultaneous visualization of sheared phospholipid monolayers.}}}\ }\href
  {https://doi.org/10.1038/ncomms1321} {\bibfield  {journal} {\bibinfo
  {journal} {Nature communications}\ }\textbf {\bibinfo {volume} {2}},\
  \bibinfo {pages} {312} (\bibinfo {year} {2011})}\BibitemShut {NoStop}%
\bibitem [{\citenamefont {Espinosa}\ \emph {et~al.}(2011)\citenamefont
  {Espinosa}, \citenamefont {L{\'{o}}pez-Montero}, \citenamefont {Monroya},\
  and\ \citenamefont {Langevin}}]{Espinosa2011}%
  \BibitemOpen
  \bibfield  {author} {\bibinfo {author} {\bibfnamefont {G.}~\bibnamefont
  {Espinosa}}, \bibinfo {author} {\bibfnamefont {I.}~\bibnamefont
  {L{\'{o}}pez-Montero}}, \bibinfo {author} {\bibfnamefont {F.}~\bibnamefont
  {Monroya}},\ and\ \bibinfo {author} {\bibfnamefont {D.}~\bibnamefont
  {Langevin}},\ }\bibfield  {title} {\enquote {\bibinfo {title} {{Shear
  rheology of lipid monolayers and insights on membrane fluidity}},}\ }\href
  {https://doi.org/10.1073/pnas.1018572108} {\bibfield  {journal} {\bibinfo
  {journal} {Proceedings of the National Academy of Sciences of the United
  States of America}\ }\textbf {\bibinfo {volume} {108}},\ \bibinfo {pages}
  {6008--6013} (\bibinfo {year} {2011})}\BibitemShut {NoStop}%
\bibitem [{\citenamefont {Samaniuk}\ and\ \citenamefont
  {Vermant}(2014)}]{Samaniuk2014}%
  \BibitemOpen
  \bibfield  {author} {\bibinfo {author} {\bibfnamefont {J.~R.}\ \bibnamefont
  {Samaniuk}}\ and\ \bibinfo {author} {\bibfnamefont {J.}~\bibnamefont
  {Vermant}},\ }\bibfield  {title} {\enquote {\bibinfo {title} {Micro and
  macrorheology at fluid--fluid interfaces},}\ }\href@noop {} {\bibfield
  {journal} {\bibinfo  {journal} {Soft Matter}\ }\textbf {\bibinfo {volume}
  {10}},\ \bibinfo {pages} {7023--7033} (\bibinfo {year} {2014})}\BibitemShut
  {NoStop}%
\bibitem [{\citenamefont {Gambin}\ \emph {et~al.}(2006)\citenamefont {Gambin},
  \citenamefont {Lopez-Esparza}, \citenamefont {Reffay}, \citenamefont
  {Sierecki}, \citenamefont {Gov}, \citenamefont {Genest}, \citenamefont
  {Hodges},\ and\ \citenamefont {Urbach}}]{Gambin2006}%
  \BibitemOpen
  \bibfield  {author} {\bibinfo {author} {\bibfnamefont {Y.}~\bibnamefont
  {Gambin}}, \bibinfo {author} {\bibfnamefont {R.}~\bibnamefont
  {Lopez-Esparza}}, \bibinfo {author} {\bibfnamefont {M.}~\bibnamefont
  {Reffay}}, \bibinfo {author} {\bibfnamefont {E.}~\bibnamefont {Sierecki}},
  \bibinfo {author} {\bibfnamefont {N.~S.}\ \bibnamefont {Gov}}, \bibinfo
  {author} {\bibfnamefont {M.}~\bibnamefont {Genest}}, \bibinfo {author}
  {\bibfnamefont {R.~S.}\ \bibnamefont {Hodges}},\ and\ \bibinfo {author}
  {\bibfnamefont {W.}~\bibnamefont {Urbach}},\ }\bibfield  {title} {\enquote
  {\bibinfo {title} {{Lateral mobility of proteins in liquid membranes
  revisited}},}\ }\href {https://doi.org/10.1073/pnas.0511026103} {\bibfield
  {journal} {\bibinfo  {journal} {Proceedings of the National Academy of
  Sciences}\ }\textbf {\bibinfo {volume} {103}},\ \bibinfo {pages} {2098--2102}
  (\bibinfo {year} {2006})}\BibitemShut {NoStop}%
\bibitem [{\citenamefont {Cicuta}, \citenamefont {Keller},\ and\ \citenamefont
  {Veatch}(2007)}]{Cicuta2007}%
  \BibitemOpen
  \bibfield  {author} {\bibinfo {author} {\bibfnamefont {P.}~\bibnamefont
  {Cicuta}}, \bibinfo {author} {\bibfnamefont {S.~L.}\ \bibnamefont {Keller}},\
  and\ \bibinfo {author} {\bibfnamefont {S.~L.}\ \bibnamefont {Veatch}},\
  }\bibfield  {title} {\enquote {\bibinfo {title} {{Diffusion of liquid domains
  in lipid bilayer membranes}},}\ }\href {https://doi.org/10.1021/jp0702088}
  {\bibfield  {journal} {\bibinfo  {journal} {Journal of Physical Chemistry B}\
  }\textbf {\bibinfo {volume} {111}},\ \bibinfo {pages} {3328--3331} (\bibinfo
  {year} {2007})}\BibitemShut {NoStop}%
\bibitem [{\citenamefont {Sch{\"u}tte}\ \emph {et~al.}(2017)\citenamefont
  {Sch{\"u}tte}, \citenamefont {Mey}, \citenamefont {Enderlein}, \citenamefont
  {Savi{\'c}}, \citenamefont {Geil}, \citenamefont {Janshoff},\ and\
  \citenamefont {Steinem}}]{Schutte2017}%
  \BibitemOpen
  \bibfield  {author} {\bibinfo {author} {\bibfnamefont {O.~M.}\ \bibnamefont
  {Sch{\"u}tte}}, \bibinfo {author} {\bibfnamefont {I.}~\bibnamefont {Mey}},
  \bibinfo {author} {\bibfnamefont {J.}~\bibnamefont {Enderlein}}, \bibinfo
  {author} {\bibfnamefont {F.}~\bibnamefont {Savi{\'c}}}, \bibinfo {author}
  {\bibfnamefont {B.}~\bibnamefont {Geil}}, \bibinfo {author} {\bibfnamefont
  {A.}~\bibnamefont {Janshoff}},\ and\ \bibinfo {author} {\bibfnamefont
  {C.}~\bibnamefont {Steinem}},\ }\bibfield  {title} {\enquote {\bibinfo
  {title} {Size and mobility of lipid domains tuned by geometrical
  constraints},}\ }\href@noop {} {\bibfield  {journal} {\bibinfo  {journal}
  {Proceedings of the National Academy of Sciences}\ }\textbf {\bibinfo
  {volume} {114}},\ \bibinfo {pages} {E6064--E6071} (\bibinfo {year}
  {2017})}\BibitemShut {NoStop}%
\bibitem [{\citenamefont {Prasad}, \citenamefont {Koehler},\ and\ \citenamefont
  {Weeks}(2006)}]{Prasad2006}%
  \BibitemOpen
  \bibfield  {author} {\bibinfo {author} {\bibfnamefont {V.}~\bibnamefont
  {Prasad}}, \bibinfo {author} {\bibfnamefont {S.}~\bibnamefont {Koehler}},\
  and\ \bibinfo {author} {\bibfnamefont {E.~R.}\ \bibnamefont {Weeks}},\
  }\bibfield  {title} {\enquote {\bibinfo {title} {Two-particle microrheology
  of quasi-2d viscous systems},}\ }\href@noop {} {\bibfield  {journal}
  {\bibinfo  {journal} {Physical Review Letters}\ }\textbf {\bibinfo {volume}
  {97}},\ \bibinfo {pages} {176001} (\bibinfo {year} {2006})}\BibitemShut
  {NoStop}%
\bibitem [{\citenamefont {Pepicelli}\ \emph {et~al.}(2019)\citenamefont
  {Pepicelli}, \citenamefont {Jaensson}, \citenamefont {Tregou{\"{e}}t},
  \citenamefont {Schroyen}, \citenamefont {Alicke}, \citenamefont {Tervoort},
  \citenamefont {Monteux},\ and\ \citenamefont {Vermant}}]{Pepicelli2019}%
  \BibitemOpen
  \bibfield  {author} {\bibinfo {author} {\bibfnamefont {M.}~\bibnamefont
  {Pepicelli}}, \bibinfo {author} {\bibfnamefont {N.}~\bibnamefont {Jaensson}},
  \bibinfo {author} {\bibfnamefont {C.}~\bibnamefont {Tregou{\"{e}}t}},
  \bibinfo {author} {\bibfnamefont {B.}~\bibnamefont {Schroyen}}, \bibinfo
  {author} {\bibfnamefont {A.}~\bibnamefont {Alicke}}, \bibinfo {author}
  {\bibfnamefont {T.}~\bibnamefont {Tervoort}}, \bibinfo {author}
  {\bibfnamefont {C.}~\bibnamefont {Monteux}},\ and\ \bibinfo {author}
  {\bibfnamefont {J.}~\bibnamefont {Vermant}},\ }\bibfield  {title} {\enquote
  {\bibinfo {title} {{Surface viscoelasticity in model polymer multilayers:
  From planar interfaces to rising bubbles}},}\ }\href
  {https://doi.org/10.1122/1.5096887} {\bibfield  {journal} {\bibinfo
  {journal} {Journal of Rheology}\ }\textbf {\bibinfo {volume} {63}},\ \bibinfo
  {pages} {815--828} (\bibinfo {year} {2019})}\BibitemShut {NoStop}%
\bibitem [{\citenamefont {Stone}\ and\ \citenamefont
  {Ajdari}(1998)}]{Stone1998}%
  \BibitemOpen
  \bibfield  {author} {\bibinfo {author} {\bibfnamefont {H.~A.}\ \bibnamefont
  {Stone}}\ and\ \bibinfo {author} {\bibfnamefont {A.}~\bibnamefont {Ajdari}},\
  }\bibfield  {title} {\enquote {\bibinfo {title} {{Hydrodynamics of particles
  embedded in a flat surfactant layer overlying a subphase of finite depth}},}\
  }\href
  {http://journals.cambridge.org/action/displayAbstract?fromPage=online&aid=14409}
  {\bibfield  {journal} {\bibinfo  {journal} {Journal of Fluid Mechanics}\
  }\textbf {\bibinfo {volume} {369}},\ \bibinfo {pages} {151--173} (\bibinfo
  {year} {1998})}\BibitemShut {NoStop}%
\bibitem [{\citenamefont {Oppenheimer}\ and\ \citenamefont
  {Diamant}(2009)}]{Oppenheimer2009}%
  \BibitemOpen
  \bibfield  {author} {\bibinfo {author} {\bibfnamefont {N.}~\bibnamefont
  {Oppenheimer}}\ and\ \bibinfo {author} {\bibfnamefont {H.}~\bibnamefont
  {Diamant}},\ }\bibfield  {title} {\enquote {\bibinfo {title} {{Correlated
  Diffusion of Membrane Proteins and Their Effect on Membrane Viscosity}},}\
  }\href {https://doi.org/10.1016/j.bpj.2009.01.020} {\bibfield  {journal}
  {\bibinfo  {journal} {Biophysical Journal}\ }\textbf {\bibinfo {volume}
  {96}},\ \bibinfo {pages} {3041--3049} (\bibinfo {year} {2009})}\BibitemShut
  {NoStop}%
\bibitem [{\citenamefont {Camley}\ and\ \citenamefont
  {Brown}(2013)}]{Camley2013}%
  \BibitemOpen
  \bibfield  {author} {\bibinfo {author} {\bibfnamefont {B.~A.}\ \bibnamefont
  {Camley}}\ and\ \bibinfo {author} {\bibfnamefont {F.~L.}\ \bibnamefont
  {Brown}},\ }\bibfield  {title} {\enquote {\bibinfo {title} {Diffusion of
  complex objects embedded in free and supported lipid bilayer membranes: role
  of shape anisotropy and leaflet structure},}\ }\href@noop {} {\bibfield
  {journal} {\bibinfo  {journal} {Soft Matter}\ }\textbf {\bibinfo {volume}
  {9}},\ \bibinfo {pages} {4767--4779} (\bibinfo {year} {2013})}\BibitemShut
  {NoStop}%
\bibitem [{\citenamefont {Fischer}(2004{\natexlab{a}})}]{Fischer2004}%
  \BibitemOpen
  \bibfield  {author} {\bibinfo {author} {\bibfnamefont {T.~M.}\ \bibnamefont
  {Fischer}},\ }\bibfield  {title} {\enquote {\bibinfo {title} {{The drag on
  needles moving in a Langmuir monolayer}},}\ }\href
  {https://doi.org/10.1017/S0022112003006608} {\bibfield  {journal} {\bibinfo
  {journal} {Journal of Fluid Mechanics}\ }\textbf {\bibinfo {volume} {498}},\
  \bibinfo {pages} {123--137} (\bibinfo {year}
  {2004}{\natexlab{a}})}\BibitemShut {NoStop}%
\bibitem [{\citenamefont {Levine}, \citenamefont {Liverpool},\ and\
  \citenamefont {MacKintosh}(2004)}]{Levine2004}%
  \BibitemOpen
  \bibfield  {author} {\bibinfo {author} {\bibfnamefont {A.~J.}\ \bibnamefont
  {Levine}}, \bibinfo {author} {\bibfnamefont {T.}~\bibnamefont {Liverpool}},\
  and\ \bibinfo {author} {\bibfnamefont {F.~C.}\ \bibnamefont {MacKintosh}},\
  }\bibfield  {title} {\enquote {\bibinfo {title} {Dynamics of rigid and
  flexible extended bodies in viscous films and membranes},}\ }\href@noop {}
  {\bibfield  {journal} {\bibinfo  {journal} {Physical review letters}\
  }\textbf {\bibinfo {volume} {93}},\ \bibinfo {pages} {038102} (\bibinfo
  {year} {2004})}\BibitemShut {NoStop}%
\bibitem [{\citenamefont {Shi}, \citenamefont {Moradi},\ and\ \citenamefont
  {Nazockdast}(2022)}]{Shi2022}%
  \BibitemOpen
  \bibfield  {author} {\bibinfo {author} {\bibfnamefont {W.}~\bibnamefont
  {Shi}}, \bibinfo {author} {\bibfnamefont {M.}~\bibnamefont {Moradi}},\ and\
  \bibinfo {author} {\bibfnamefont {E.}~\bibnamefont {Nazockdast}},\ }\bibfield
   {title} {\enquote {\bibinfo {title} {Hydrodynamics of a single filament
  moving in a spherical membrane},}\ }\href@noop {} {\bibfield  {journal}
  {\bibinfo  {journal} {Physical Review Fluids}\ }\textbf {\bibinfo {volume}
  {7}},\ \bibinfo {pages} {084004} (\bibinfo {year} {2022})}\BibitemShut
  {NoStop}%
\bibitem [{\citenamefont {Mikhailov}\ and\ \citenamefont
  {Kapral}(2015)}]{actprot}%
  \BibitemOpen
  \bibfield  {author} {\bibinfo {author} {\bibfnamefont {A.~S.}\ \bibnamefont
  {Mikhailov}}\ and\ \bibinfo {author} {\bibfnamefont {R.}~\bibnamefont
  {Kapral}},\ }\bibfield  {title} {\enquote {\bibinfo {title} {{Hydrodynamic
  collective effects of active protein machines in solution and lipid
  bilayers}},}\ }\href {https://doi.org/10.1073/pnas.1506825112} {\bibfield
  {journal} {\bibinfo  {journal} {Proceedings of the National Academy of
  Sciences}\ }\textbf {\bibinfo {volume} {112}},\ \bibinfo {pages}
  {E3639--E3644} (\bibinfo {year} {2015})}\BibitemShut {NoStop}%
\bibitem [{\citenamefont {Manikantan}(2020)}]{Manikantan2020PRL}%
  \BibitemOpen
  \bibfield  {author} {\bibinfo {author} {\bibfnamefont {H.}~\bibnamefont
  {Manikantan}},\ }\bibfield  {title} {\enquote {\bibinfo {title} {{Tunable
  Collective Dynamics of Active Inclusions in Viscous Membranes}},}\ }\href
  {https://doi.org/10.1103/PhysRevLett.125.268101} {\bibfield  {journal}
  {\bibinfo  {journal} {Physical Review Letters}\ }\textbf {\bibinfo {volume}
  {125}},\ \bibinfo {pages} {268101} (\bibinfo {year} {2020})}\BibitemShut
  {NoStop}%
\bibitem [{\citenamefont {Oppenheimer}, \citenamefont {Stein},\ and\
  \citenamefont {Shelley}(2019)}]{Oppenheimer2019}%
  \BibitemOpen
  \bibfield  {author} {\bibinfo {author} {\bibfnamefont {N.}~\bibnamefont
  {Oppenheimer}}, \bibinfo {author} {\bibfnamefont {D.~B.}\ \bibnamefont
  {Stein}},\ and\ \bibinfo {author} {\bibfnamefont {M.~J.}\ \bibnamefont
  {Shelley}},\ }\bibfield  {title} {\enquote {\bibinfo {title} {Rotating
  membrane inclusions crystallize through hydrodynamic and steric
  interactions},}\ }\href@noop {} {\bibfield  {journal} {\bibinfo  {journal}
  {Physical Review Letters}\ }\textbf {\bibinfo {volume} {123}},\ \bibinfo
  {pages} {148101} (\bibinfo {year} {2019})}\BibitemShut {NoStop}%
\bibitem [{\citenamefont {Ramachandran}\ \emph {et~al.}(2011)\citenamefont
  {Ramachandran}, \citenamefont {Komura}, \citenamefont {Seki},\ and\
  \citenamefont {Gompper}}]{Ramachandran2011}%
  \BibitemOpen
  \bibfield  {author} {\bibinfo {author} {\bibfnamefont {S.}~\bibnamefont
  {Ramachandran}}, \bibinfo {author} {\bibfnamefont {S.}~\bibnamefont
  {Komura}}, \bibinfo {author} {\bibfnamefont {K.}~\bibnamefont {Seki}},\ and\
  \bibinfo {author} {\bibfnamefont {G.}~\bibnamefont {Gompper}},\ }\bibfield
  {title} {\enquote {\bibinfo {title} {{Dynamics of a polymer chain confined in
  a membrane}},}\ }\href {https://doi.org/10.1140/epje/i2011-11046-3}
  {\bibfield  {journal} {\bibinfo  {journal} {The European Physical Journal E}\
  }\textbf {\bibinfo {volume} {34}},\ \bibinfo {pages} {46} (\bibinfo {year}
  {2011})}\BibitemShut {NoStop}%
\bibitem [{\citenamefont {Kaganer}, \citenamefont {M\"ohwald},\ and\
  \citenamefont {Dutta}(1999)}]{Kaganer1999}%
  \BibitemOpen
  \bibfield  {author} {\bibinfo {author} {\bibfnamefont {V.~M.}\ \bibnamefont
  {Kaganer}}, \bibinfo {author} {\bibfnamefont {H.}~\bibnamefont {M\"ohwald}},\
  and\ \bibinfo {author} {\bibfnamefont {P.}~\bibnamefont {Dutta}},\ }\bibfield
   {title} {\enquote {\bibinfo {title} {Structure and phase transitions in
  langmuir monolayers},}\ }\href@noop {} {\bibfield  {journal} {\bibinfo
  {journal} {Rev. Mod. Phys.}\ }\textbf {\bibinfo {volume} {71}},\ \bibinfo
  {pages} {779--819} (\bibinfo {year} {1999})}\BibitemShut {NoStop}%
\bibitem [{\citenamefont {Lingwood}\ and\ \citenamefont
  {Simons}(2010)}]{Lingwood2010}%
  \BibitemOpen
  \bibfield  {author} {\bibinfo {author} {\bibfnamefont {D.}~\bibnamefont
  {Lingwood}}\ and\ \bibinfo {author} {\bibfnamefont {K.}~\bibnamefont
  {Simons}},\ }\bibfield  {title} {\enquote {\bibinfo {title} {{Lipid Rafts As
  a Membrane-Organizing Principle}},}\ }\href
  {https://doi.org/10.1126/science.1174621} {\bibfield  {journal} {\bibinfo
  {journal} {Science}\ }\textbf {\bibinfo {volume} {327}},\ \bibinfo {pages}
  {46--50} (\bibinfo {year} {2010})}\BibitemShut {NoStop}%
\bibitem [{\citenamefont {Sezgin}\ \emph {et~al.}(2017)\citenamefont {Sezgin},
  \citenamefont {Levental}, \citenamefont {Mayor},\ and\ \citenamefont
  {Eggeling}}]{Sezgin2017}%
  \BibitemOpen
  \bibfield  {author} {\bibinfo {author} {\bibfnamefont {E.}~\bibnamefont
  {Sezgin}}, \bibinfo {author} {\bibfnamefont {I.}~\bibnamefont {Levental}},
  \bibinfo {author} {\bibfnamefont {S.}~\bibnamefont {Mayor}},\ and\ \bibinfo
  {author} {\bibfnamefont {C.}~\bibnamefont {Eggeling}},\ }\bibfield  {title}
  {\enquote {\bibinfo {title} {{The mystery of membrane organization:
  Composition, regulation and roles of lipid rafts}},}\ }\href
  {https://doi.org/10.1038/nrm.2017.16} {\bibfield  {journal} {\bibinfo
  {journal} {Nature Reviews Molecular Cell Biology}\ }\textbf {\bibinfo
  {volume} {18}},\ \bibinfo {pages} {361--374} (\bibinfo {year}
  {2017})}\BibitemShut {NoStop}%
\bibitem [{\citenamefont {Arriaga}\ \emph {et~al.}(2010)\citenamefont
  {Arriaga}, \citenamefont {L{\'o}pez-Montero}, \citenamefont
  {Ign{\'e}s-Mullol},\ and\ \citenamefont {Monroy}}]{Arriaga2010}%
  \BibitemOpen
  \bibfield  {author} {\bibinfo {author} {\bibfnamefont {L.~R.}\ \bibnamefont
  {Arriaga}}, \bibinfo {author} {\bibfnamefont {I.}~\bibnamefont
  {L{\'o}pez-Montero}}, \bibinfo {author} {\bibfnamefont {J.}~\bibnamefont
  {Ign{\'e}s-Mullol}},\ and\ \bibinfo {author} {\bibfnamefont {F.}~\bibnamefont
  {Monroy}},\ }\bibfield  {title} {\enquote {\bibinfo {title} {Domain-growth
  kinetic origin of nonhorizontal phase coexistence plateaux in langmuir
  monolayers: compression rigidity of a raft-like lipid distribution},}\
  }\href@noop {} {\bibfield  {journal} {\bibinfo  {journal} {The Journal of
  Physical Chemistry B}\ }\textbf {\bibinfo {volume} {114}},\ \bibinfo {pages}
  {4509--4520} (\bibinfo {year} {2010})}\BibitemShut {NoStop}%
\bibitem [{\citenamefont {Kurnaz}\ and\ \citenamefont
  {Schwartz}(1997)}]{Kurnaz1997}%
  \BibitemOpen
  \bibfield  {author} {\bibinfo {author} {\bibfnamefont {M.~L.}\ \bibnamefont
  {Kurnaz}}\ and\ \bibinfo {author} {\bibfnamefont {D.~K.}\ \bibnamefont
  {Schwartz}},\ }\bibfield  {title} {\enquote {\bibinfo {title} {{Channel flow
  in a Langmuir monolayer: Unusual velocity profiles in a liquid-crystalline
  mesophase}},}\ }\href {https://doi.org/10.1103/PhysRevE.56.3378} {\bibfield
  {journal} {\bibinfo  {journal} {Physical Review E}\ }\textbf {\bibinfo
  {volume} {56}},\ \bibinfo {pages} {3378--3384} (\bibinfo {year}
  {1997})}\BibitemShut {NoStop}%
\bibitem [{\citenamefont {Kim}\ \emph {et~al.}(2011)\citenamefont {Kim},
  \citenamefont {Choi}, \citenamefont {Zasadzinski},\ and\ \citenamefont
  {Squires}}]{Kim2011}%
  \BibitemOpen
  \bibfield  {author} {\bibinfo {author} {\bibfnamefont {K.}~\bibnamefont
  {Kim}}, \bibinfo {author} {\bibfnamefont {S.~Q.}\ \bibnamefont {Choi}},
  \bibinfo {author} {\bibfnamefont {J.~A.}\ \bibnamefont {Zasadzinski}},\ and\
  \bibinfo {author} {\bibfnamefont {T.~M.}\ \bibnamefont {Squires}},\
  }\bibfield  {title} {\enquote {\bibinfo {title} {{Interfacial microrheology
  of DPPC monolayers at the air–water interface}},}\ }\href
  {https://doi.org/10.1039/c1sm05383c} {\bibfield  {journal} {\bibinfo
  {journal} {Soft Matter}\ }\textbf {\bibinfo {volume} {7}},\ \bibinfo {pages}
  {7782} (\bibinfo {year} {2011})}\BibitemShut {NoStop}%
\bibitem [{\citenamefont {Fogelson}\ and\ \citenamefont
  {Mogilner}(2014)}]{Fogelson2014}%
  \BibitemOpen
  \bibfield  {author} {\bibinfo {author} {\bibfnamefont {B.}~\bibnamefont
  {Fogelson}}\ and\ \bibinfo {author} {\bibfnamefont {A.}~\bibnamefont
  {Mogilner}},\ }\bibfield  {title} {\enquote {\bibinfo {title} {Computational
  estimates of membrane flow and tension gradient in motile cells},}\
  }\href@noop {} {\bibfield  {journal} {\bibinfo  {journal} {PloS One}\
  }\textbf {\bibinfo {volume} {9}},\ \bibinfo {pages} {e84524} (\bibinfo {year}
  {2014})}\BibitemShut {NoStop}%
\bibitem [{\citenamefont {Bruinsma}\ and\ \citenamefont
  {Pincus}(1996)}]{Bruinsma1996}%
  \BibitemOpen
  \bibfield  {author} {\bibinfo {author} {\bibfnamefont {R.}~\bibnamefont
  {Bruinsma}}\ and\ \bibinfo {author} {\bibfnamefont {P.}~\bibnamefont
  {Pincus}},\ }\bibfield  {title} {\enquote {\bibinfo {title} {Protein
  aggregation in membranes},}\ }\href@noop {} {\bibfield  {journal} {\bibinfo
  {journal} {Current Opinion in Solid State and Materials Science}\ }\textbf
  {\bibinfo {volume} {1}},\ \bibinfo {pages} {401--406} (\bibinfo {year}
  {1996})}\BibitemShut {NoStop}%
\bibitem [{\citenamefont {Manikantan}\ and\ \citenamefont
  {Squires}(2017{\natexlab{a}})}]{HMPRF}%
  \BibitemOpen
  \bibfield  {author} {\bibinfo {author} {\bibfnamefont {H.}~\bibnamefont
  {Manikantan}}\ and\ \bibinfo {author} {\bibfnamefont {T.~M.}\ \bibnamefont
  {Squires}},\ }\bibfield  {title} {\enquote {\bibinfo {title}
  {{Pressure-dependent surface viscosity and its surprising consequences in
  interfacial lubrication flows}},}\ }\href
  {https://doi.org/10.1103/PhysRevFluids.2.023301} {\bibfield  {journal}
  {\bibinfo  {journal} {Physical Review Fluids}\ }\textbf {\bibinfo {volume}
  {2}},\ \bibinfo {pages} {023301} (\bibinfo {year}
  {2017}{\natexlab{a}})}\BibitemShut {NoStop}%
\bibitem [{\citenamefont {Manikantan}\ and\ \citenamefont
  {Squires}(2017{\natexlab{b}})}]{HMPRSA}%
  \BibitemOpen
  \bibfield  {author} {\bibinfo {author} {\bibfnamefont {H.}~\bibnamefont
  {Manikantan}}\ and\ \bibinfo {author} {\bibfnamefont {T.~M.}\ \bibnamefont
  {Squires}},\ }\bibfield  {title} {\enquote {\bibinfo {title} {{Irreversible
  particle motion in surfactant-laden interfaces due to pressure-dependent
  surface viscosity}},}\ }\href {https://doi.org/10.1098/rspa.2017.0346}
  {\bibfield  {journal} {\bibinfo  {journal} {Proceedings of the Royal Society
  A: Mathematical, Physical and Engineering Science}\ }\textbf {\bibinfo
  {volume} {473}},\ \bibinfo {pages} {20170346} (\bibinfo {year}
  {2017}{\natexlab{b}})}\BibitemShut {NoStop}%
\bibitem [{\citenamefont {Singh}\ and\ \citenamefont
  {Narsimhan}(2021)}]{Singh2021}%
  \BibitemOpen
  \bibfield  {author} {\bibinfo {author} {\bibfnamefont {N.}~\bibnamefont
  {Singh}}\ and\ \bibinfo {author} {\bibfnamefont {V.}~\bibnamefont
  {Narsimhan}},\ }\bibfield  {title} {\enquote {\bibinfo {title} {Impact of
  surface viscosity on the stability of a droplet translating through a
  stagnant fluid},}\ }\href@noop {} {\bibfield  {journal} {\bibinfo  {journal}
  {Journal of Fluid Mechanics}\ }\textbf {\bibinfo {volume} {927}},\ \bibinfo
  {pages} {A44} (\bibinfo {year} {2021})}\BibitemShut {NoStop}%
\bibitem [{\citenamefont {Herrada}\ \emph {et~al.}(2022)\citenamefont
  {Herrada}, \citenamefont {Ponce-Torres}, \citenamefont {Rubio}, \citenamefont
  {Eggers},\ and\ \citenamefont {Montanero}}]{Herrada2022}%
  \BibitemOpen
  \bibfield  {author} {\bibinfo {author} {\bibfnamefont {M.~A.}\ \bibnamefont
  {Herrada}}, \bibinfo {author} {\bibfnamefont {A.}~\bibnamefont
  {Ponce-Torres}}, \bibinfo {author} {\bibfnamefont {M.}~\bibnamefont {Rubio}},
  \bibinfo {author} {\bibfnamefont {J.}~\bibnamefont {Eggers}},\ and\ \bibinfo
  {author} {\bibfnamefont {J.~M.}\ \bibnamefont {Montanero}},\ }\bibfield
  {title} {\enquote {\bibinfo {title} {Stability and tip streaming of a
  surfactant-loaded drop in an extensional flow. influence of surface
  viscosity},}\ }\href@noop {} {\bibfield  {journal} {\bibinfo  {journal}
  {Journal of Fluid Mechanics}\ }\textbf {\bibinfo {volume} {934}},\ \bibinfo
  {pages} {A26} (\bibinfo {year} {2022})}\BibitemShut {NoStop}%
\bibitem [{\citenamefont {Reynwar}\ \emph {et~al.}(2007)\citenamefont
  {Reynwar}, \citenamefont {Illya}, \citenamefont {Harmandaris}, \citenamefont
  {M{\"u}ller}, \citenamefont {Kremer},\ and\ \citenamefont
  {Deserno}}]{Reynwar2007}%
  \BibitemOpen
  \bibfield  {author} {\bibinfo {author} {\bibfnamefont {B.~J.}\ \bibnamefont
  {Reynwar}}, \bibinfo {author} {\bibfnamefont {G.}~\bibnamefont {Illya}},
  \bibinfo {author} {\bibfnamefont {V.~A.}\ \bibnamefont {Harmandaris}},
  \bibinfo {author} {\bibfnamefont {M.~M.}\ \bibnamefont {M{\"u}ller}},
  \bibinfo {author} {\bibfnamefont {K.}~\bibnamefont {Kremer}},\ and\ \bibinfo
  {author} {\bibfnamefont {M.}~\bibnamefont {Deserno}},\ }\bibfield  {title}
  {\enquote {\bibinfo {title} {Aggregation and vesiculation of membrane
  proteins by curvature-mediated interactions},}\ }\href@noop {} {\bibfield
  {journal} {\bibinfo  {journal} {Nature}\ }\textbf {\bibinfo {volume} {447}},\
  \bibinfo {pages} {461--464} (\bibinfo {year} {2007})}\BibitemShut {NoStop}%
\bibitem [{\citenamefont {Dasgupta}, \citenamefont {Auth},\ and\ \citenamefont
  {Gompper}(2017)}]{Dasgupta2017}%
  \BibitemOpen
  \bibfield  {author} {\bibinfo {author} {\bibfnamefont {S.}~\bibnamefont
  {Dasgupta}}, \bibinfo {author} {\bibfnamefont {T.}~\bibnamefont {Auth}},\
  and\ \bibinfo {author} {\bibfnamefont {G.}~\bibnamefont {Gompper}},\
  }\bibfield  {title} {\enquote {\bibinfo {title} {{Nano- and microparticles at
  fluid and biological interfaces}},}\ }\href
  {https://doi.org/10.1088/1361-648X/aa7933} {\bibfield  {journal} {\bibinfo
  {journal} {Journal of Physics: Condensed Matter}\ }\textbf {\bibinfo {volume}
  {29}},\ \bibinfo {pages} {373003} (\bibinfo {year} {2017})}\BibitemShut
  {NoStop}%
\bibitem [{\citenamefont {Slattery}\ and\ \citenamefont {Sagis}(2007)}]{Slate}%
  \BibitemOpen
  \bibfield  {author} {\bibinfo {author} {\bibfnamefont {J.~C.}\ \bibnamefont
  {Slattery}}\ and\ \bibinfo {author} {\bibfnamefont {L.}~\bibnamefont
  {Sagis}},\ }\href@noop {} {\emph {\bibinfo {title} {Interfacial Transport
  Phenomena.}}}\ (\bibinfo  {publisher} {Springer},\ \bibinfo {year}
  {2007})\BibitemShut {NoStop}%
\bibitem [{\citenamefont {Manikantan}\ and\ \citenamefont
  {Squires}(2020)}]{Manikantan2020}%
  \BibitemOpen
  \bibfield  {author} {\bibinfo {author} {\bibfnamefont {H.}~\bibnamefont
  {Manikantan}}\ and\ \bibinfo {author} {\bibfnamefont {T.~M.}\ \bibnamefont
  {Squires}},\ }\bibfield  {title} {\enquote {\bibinfo {title} {{Surfactant
  dynamics: hidden variables controlling fluid flows}},}\ }\href
  {https://doi.org/10.1017/jfm.2020.170} {\bibfield  {journal} {\bibinfo
  {journal} {Journal of Fluid Mechanics}\ }\textbf {\bibinfo {volume} {892}},\
  \bibinfo {pages} {P1} (\bibinfo {year} {2020})}\BibitemShut {NoStop}%
\bibitem [{\citenamefont {Fischer}(2004{\natexlab{b}})}]{Fischer_comment}%
  \BibitemOpen
  \bibfield  {author} {\bibinfo {author} {\bibfnamefont {T.~M.}\ \bibnamefont
  {Fischer}},\ }\bibfield  {title} {\enquote {\bibinfo {title} {Comment on
  ``shear viscosity of langmuir monolayers in the low-density limit''},}\
  }\href {https://doi.org/10.1103/PhysRevLett.92.139603} {\bibfield  {journal}
  {\bibinfo  {journal} {Physical Review Letters}\ }\textbf {\bibinfo {volume}
  {92}},\ \bibinfo {pages} {139603} (\bibinfo {year}
  {2004}{\natexlab{b}})}\BibitemShut {NoStop}%
\bibitem [{\citenamefont {Stone}\ and\ \citenamefont
  {Masoud}(2015)}]{Stone2015}%
  \BibitemOpen
  \bibfield  {author} {\bibinfo {author} {\bibfnamefont {H.~A.}\ \bibnamefont
  {Stone}}\ and\ \bibinfo {author} {\bibfnamefont {H.}~\bibnamefont {Masoud}},\
  }\bibfield  {title} {\enquote {\bibinfo {title} {{Mobility of
  membrane-trapped particles}},}\ }\href {https://doi.org/10.1017/jfm.2015.486}
  {\bibfield  {journal} {\bibinfo  {journal} {Journal of Fluid Mechanics}\
  }\textbf {\bibinfo {volume} {781}},\ \bibinfo {pages} {494--505} (\bibinfo
  {year} {2015})}\BibitemShut {NoStop}%
\bibitem [{\citenamefont {Alonso}\ and\ \citenamefont
  {Zasadzinski}(2006)}]{Alonso2006}%
  \BibitemOpen
  \bibfield  {author} {\bibinfo {author} {\bibfnamefont {C.}~\bibnamefont
  {Alonso}}\ and\ \bibinfo {author} {\bibfnamefont {J.~A.}\ \bibnamefont
  {Zasadzinski}},\ }\bibfield  {title} {\enquote {\bibinfo {title} {A brief
  review of the relationships between monolayer viscosity, phase behavior,
  surface pressure, and temperature using a simple monolayer viscometer},}\
  }\href@noop {} {\bibfield  {journal} {\bibinfo  {journal} {The Journal of
  Physical Chemistry B}\ }\textbf {\bibinfo {volume} {110}},\ \bibinfo {pages}
  {22185--22191} (\bibinfo {year} {2006})}\BibitemShut {NoStop}%
\bibitem [{\citenamefont {Zell}\ \emph {et~al.}(2014)\citenamefont {Zell},
  \citenamefont {Nowbahar}, \citenamefont {Mansard}, \citenamefont {Leal},
  \citenamefont {Deshmukh}, \citenamefont {Mecca}, \citenamefont {Tucker},\
  and\ \citenamefont {Squires}}]{Zell2014}%
  \BibitemOpen
  \bibfield  {author} {\bibinfo {author} {\bibfnamefont {Z.~A.}\ \bibnamefont
  {Zell}}, \bibinfo {author} {\bibfnamefont {A.}~\bibnamefont {Nowbahar}},
  \bibinfo {author} {\bibfnamefont {V.}~\bibnamefont {Mansard}}, \bibinfo
  {author} {\bibfnamefont {L.~G.}\ \bibnamefont {Leal}}, \bibinfo {author}
  {\bibfnamefont {S.~S.}\ \bibnamefont {Deshmukh}}, \bibinfo {author}
  {\bibfnamefont {J.~M.}\ \bibnamefont {Mecca}}, \bibinfo {author}
  {\bibfnamefont {C.~J.}\ \bibnamefont {Tucker}},\ and\ \bibinfo {author}
  {\bibfnamefont {T.~M.}\ \bibnamefont {Squires}},\ }\bibfield  {title}
  {\enquote {\bibinfo {title} {{Surface shear inviscidity of soluble
  surfactants}},}\ }\href {https://doi.org/10.1073/pnas.1315991111} {\bibfield
  {journal} {\bibinfo  {journal} {Proceedings of the National Academy of
  Sciences}\ }\textbf {\bibinfo {volume} {111}},\ \bibinfo {pages} {3677--3682}
  (\bibinfo {year} {2014})}\BibitemShut {NoStop}%
\bibitem [{\citenamefont {Kim}\ \emph {et~al.}(2013)\citenamefont {Kim},
  \citenamefont {Choi}, \citenamefont {Zell}, \citenamefont {Squires},\ and\
  \citenamefont {Zasadzinski}}]{Kim2013}%
  \BibitemOpen
  \bibfield  {author} {\bibinfo {author} {\bibfnamefont {K.}~\bibnamefont
  {Kim}}, \bibinfo {author} {\bibfnamefont {S.~Q.}\ \bibnamefont {Choi}},
  \bibinfo {author} {\bibfnamefont {Z.~A.}\ \bibnamefont {Zell}}, \bibinfo
  {author} {\bibfnamefont {T.~M.}\ \bibnamefont {Squires}},\ and\ \bibinfo
  {author} {\bibfnamefont {J.~A.}\ \bibnamefont {Zasadzinski}},\ }\bibfield
  {title} {\enquote {\bibinfo {title} {{Effect of cholesterol nanodomains on
  monolayer morphology and dynamics}},}\ }\href
  {https://doi.org/10.1073/pnas.1303304110} {\bibfield  {journal} {\bibinfo
  {journal} {Proceedings of the National Academy of Sciences}\ }\textbf
  {\bibinfo {volume} {110}},\ \bibinfo {pages} {E3054--E3060} (\bibinfo {year}
  {2013})}\BibitemShut {NoStop}%
\bibitem [{\citenamefont {Leal}(2007)}]{Leal}%
  \BibitemOpen
  \bibfield  {author} {\bibinfo {author} {\bibfnamefont {L.~G.}\ \bibnamefont
  {Leal}},\ }\href {https://doi.org/10.1017/CBO9780511800245} {\emph {\bibinfo
  {title} {Cambridge Series in Chemical Engineering}}}\ (\bibinfo  {publisher}
  {Cambridge University Press},\ \bibinfo {address} {Cambridge},\ \bibinfo
  {year} {2007})\BibitemShut {NoStop}%
\bibitem [{\citenamefont {Pozrikidis}(1992)}]{Pozrikidis}%
  \BibitemOpen
  \bibfield  {author} {\bibinfo {author} {\bibfnamefont {C.}~\bibnamefont
  {Pozrikidis}},\ }\href@noop {} {\emph {\bibinfo {title} {Boundary integral
  and singularity methods for linearized viscous flow}}}\ (\bibinfo
  {publisher} {Cambridge university press},\ \bibinfo {year}
  {1992})\BibitemShut {NoStop}%
\bibitem [{\citenamefont {Kim}\ and\ \citenamefont {Karrila}(1991)}]{Kim}%
  \BibitemOpen
  \bibfield  {author} {\bibinfo {author} {\bibfnamefont {S.}~\bibnamefont
  {Kim}}\ and\ \bibinfo {author} {\bibfnamefont {S.~J.}\ \bibnamefont
  {Karrila}},\ }\href@noop {} {\emph {\bibinfo {title} {{Microhydrodynamics:
  principles and selected applications}}}}\ (\bibinfo  {publisher}
  {Butterworth-Heinemann},\ \bibinfo {year} {1991})\BibitemShut {NoStop}%
\bibitem [{\citenamefont {Masoud}\ and\ \citenamefont
  {Stone}(2019)}]{Masoud2019}%
  \BibitemOpen
  \bibfield  {author} {\bibinfo {author} {\bibfnamefont {H.}~\bibnamefont
  {Masoud}}\ and\ \bibinfo {author} {\bibfnamefont {H.~A.}\ \bibnamefont
  {Stone}},\ }\bibfield  {title} {\enquote {\bibinfo {title} {The reciprocal
  theorem in fluid dynamics and transport phenomena},}\ }\href@noop {}
  {\bibfield  {journal} {\bibinfo  {journal} {Journal of Fluid Mechanics}\
  }\textbf {\bibinfo {volume} {879}},\ \bibinfo {pages} {P1} (\bibinfo {year}
  {2019})}\BibitemShut {NoStop}%
\bibitem [{\citenamefont {Vishnampet}\ and\ \citenamefont
  {Saintillan}(2012)}]{Vishnampet2012}%
  \BibitemOpen
  \bibfield  {author} {\bibinfo {author} {\bibfnamefont {R.}~\bibnamefont
  {Vishnampet}}\ and\ \bibinfo {author} {\bibfnamefont {D.}~\bibnamefont
  {Saintillan}},\ }\bibfield  {title} {\enquote {\bibinfo {title}
  {Concentration instability of sedimenting spheres in a second-order fluid},}\
  }\href@noop {} {\bibfield  {journal} {\bibinfo  {journal} {Physics of
  Fluids}\ }\textbf {\bibinfo {volume} {24}},\ \bibinfo {pages} {073302}
  (\bibinfo {year} {2012})}\BibitemShut {NoStop}%
\bibitem [{\citenamefont {Chen}\ \emph {et~al.}(2021)\citenamefont {Chen},
  \citenamefont {Demir}, \citenamefont {Gao}, \citenamefont {Young},\ and\
  \citenamefont {Pak}}]{Chen2021}%
  \BibitemOpen
  \bibfield  {author} {\bibinfo {author} {\bibfnamefont {Y.}~\bibnamefont
  {Chen}}, \bibinfo {author} {\bibfnamefont {E.}~\bibnamefont {Demir}},
  \bibinfo {author} {\bibfnamefont {W.}~\bibnamefont {Gao}}, \bibinfo {author}
  {\bibfnamefont {Y.-N.}\ \bibnamefont {Young}},\ and\ \bibinfo {author}
  {\bibfnamefont {O.~S.}\ \bibnamefont {Pak}},\ }\bibfield  {title} {\enquote
  {\bibinfo {title} {Wall-induced translation of a rotating particle in a
  shear-thinning fluid},}\ }\href@noop {} {\bibfield  {journal} {\bibinfo
  {journal} {Journal of Fluid Mechanics}\ }\textbf {\bibinfo {volume} {927}},\
  \bibinfo {pages} {R2} (\bibinfo {year} {2021})}\BibitemShut {NoStop}%
\bibitem [{\citenamefont {Koch}\ and\ \citenamefont
  {Shaqfeh}(1989)}]{Koch1989}%
  \BibitemOpen
  \bibfield  {author} {\bibinfo {author} {\bibfnamefont {D.~L.}\ \bibnamefont
  {Koch}}\ and\ \bibinfo {author} {\bibfnamefont {E.~S.}\ \bibnamefont
  {Shaqfeh}},\ }\bibfield  {title} {\enquote {\bibinfo {title} {The instability
  of a dispersion of sedimenting spheroids},}\ }\href@noop {} {\bibfield
  {journal} {\bibinfo  {journal} {Journal of Fluid Mechanics}\ }\textbf
  {\bibinfo {volume} {209}},\ \bibinfo {pages} {521--542} (\bibinfo {year}
  {1989})}\BibitemShut {NoStop}%
\bibitem [{\citenamefont {Saintillan}\ and\ \citenamefont
  {Shelley}(2008)}]{Saintillan2008}%
  \BibitemOpen
  \bibfield  {author} {\bibinfo {author} {\bibfnamefont {D.}~\bibnamefont
  {Saintillan}}\ and\ \bibinfo {author} {\bibfnamefont {M.~J.}\ \bibnamefont
  {Shelley}},\ }\bibfield  {title} {\enquote {\bibinfo {title} {Instabilities
  and pattern formation in active particle suspensions: kinetic theory and
  continuum simulations},}\ }\href@noop {} {\bibfield  {journal} {\bibinfo
  {journal} {Physical Review Letters}\ }\textbf {\bibinfo {volume} {100}},\
  \bibinfo {pages} {178103} (\bibinfo {year} {2008})}\BibitemShut {NoStop}%
\bibitem [{\citenamefont {Manikantan}\ \emph {et~al.}(2014)\citenamefont
  {Manikantan}, \citenamefont {Li}, \citenamefont {Spagnolie},\ and\
  \citenamefont {Saintillan}}]{Manikantan2014}%
  \BibitemOpen
  \bibfield  {author} {\bibinfo {author} {\bibfnamefont {H.}~\bibnamefont
  {Manikantan}}, \bibinfo {author} {\bibfnamefont {L.}~\bibnamefont {Li}},
  \bibinfo {author} {\bibfnamefont {S.~E.}\ \bibnamefont {Spagnolie}},\ and\
  \bibinfo {author} {\bibfnamefont {D.}~\bibnamefont {Saintillan}},\ }\bibfield
   {title} {\enquote {\bibinfo {title} {{The instability of a sedimenting
  suspension of weakly flexible fibres}},}\ }\href
  {https://doi.org/10.1017/jfm.2014.482} {\bibfield  {journal} {\bibinfo
  {journal} {Journal of Fluid Mechanics}\ }\textbf {\bibinfo {volume} {756}},\
  \bibinfo {pages} {935--964} (\bibinfo {year} {2014})}\BibitemShut {NoStop}%
\bibitem [{\citenamefont {Hasimoto}(1959)}]{Hasimoto1959}%
  \BibitemOpen
  \bibfield  {author} {\bibinfo {author} {\bibfnamefont {H.}~\bibnamefont
  {Hasimoto}},\ }\bibfield  {title} {\enquote {\bibinfo {title} {On the
  periodic fundamental solutions of the stokes equations and their application
  to viscous flow past a cubic array of spheres},}\ }\href@noop {} {\bibfield
  {journal} {\bibinfo  {journal} {Journal of Fluid Mechanics}\ }\textbf
  {\bibinfo {volume} {5}},\ \bibinfo {pages} {317--328} (\bibinfo {year}
  {1959})}\BibitemShut {NoStop}%
\bibitem [{\citenamefont {Grzybowski}, \citenamefont {Stone},\ and\
  \citenamefont {Whitesides}(2000)}]{Grzybowski2000}%
  \BibitemOpen
  \bibfield  {author} {\bibinfo {author} {\bibfnamefont {B.~A.}\ \bibnamefont
  {Grzybowski}}, \bibinfo {author} {\bibfnamefont {H.~A.}\ \bibnamefont
  {Stone}},\ and\ \bibinfo {author} {\bibfnamefont {G.~M.}\ \bibnamefont
  {Whitesides}},\ }\bibfield  {title} {\enquote {\bibinfo {title} {Dynamic
  self-assembly of magnetized, millimetre-sized objects rotating at a
  liquid--air interface},}\ }\href {https://doi.org/10.1038/35016528}
  {\bibfield  {journal} {\bibinfo  {journal} {Nature}\ }\textbf {\bibinfo
  {volume} {405}},\ \bibinfo {pages} {1033--1036} (\bibinfo {year}
  {2000})}\BibitemShut {NoStop}%
\bibitem [{\citenamefont {Goto}\ and\ \citenamefont {Tanaka}(2015)}]{Goto2015}%
  \BibitemOpen
  \bibfield  {author} {\bibinfo {author} {\bibfnamefont {Y.}~\bibnamefont
  {Goto}}\ and\ \bibinfo {author} {\bibfnamefont {H.}~\bibnamefont {Tanaka}},\
  }\bibfield  {title} {\enquote {\bibinfo {title} {Purely hydrodynamic ordering
  of rotating disks at a finite reynolds number},}\ }\href@noop {} {\bibfield
  {journal} {\bibinfo  {journal} {Nature communications}\ }\textbf {\bibinfo
  {volume} {6}},\ \bibinfo {pages} {5994} (\bibinfo {year} {2015})}\BibitemShut
  {NoStop}%
\bibitem [{\citenamefont {Cohen}\ and\ \citenamefont {Shi}(2020)}]{Cohen2020}%
  \BibitemOpen
  \bibfield  {author} {\bibinfo {author} {\bibfnamefont {A.~E.}\ \bibnamefont
  {Cohen}}\ and\ \bibinfo {author} {\bibfnamefont {Z.}~\bibnamefont {Shi}},\
  }\bibfield  {title} {\enquote {\bibinfo {title} {Do cell membranes flow like
  honey or jiggle like jello?}}\ }\href@noop {} {\bibfield  {journal} {\bibinfo
   {journal} {BioEssays}\ }\textbf {\bibinfo {volume} {42}},\ \bibinfo {pages}
  {1900142} (\bibinfo {year} {2020})}\BibitemShut {NoStop}%
\bibitem [{\citenamefont {Hron}, \citenamefont {Malek},\ and\ \citenamefont
  {Rajagopal}(2001)}]{Hron2001}%
  \BibitemOpen
  \bibfield  {author} {\bibinfo {author} {\bibfnamefont {J.}~\bibnamefont
  {Hron}}, \bibinfo {author} {\bibfnamefont {J.}~\bibnamefont {Malek}},\ and\
  \bibinfo {author} {\bibfnamefont {K.~R.}\ \bibnamefont {Rajagopal}},\
  }\bibfield  {title} {\enquote {\bibinfo {title} {{Simple flows of fluids with
  pressure-dependent viscosities}},}\ }\href
  {https://doi.org/10.1098/rspa.2000.0723} {\bibfield  {journal} {\bibinfo
  {journal} {Proceedings of the Royal Society A: Mathematical, Physical and
  Engineering Sciences}\ }\textbf {\bibinfo {volume} {457}},\ \bibinfo {pages}
  {1603--1622} (\bibinfo {year} {2001})}\BibitemShut {NoStop}%
\bibitem [{\citenamefont {Denn}(1981)}]{Denn1981}%
  \BibitemOpen
  \bibfield  {author} {\bibinfo {author} {\bibfnamefont {M.~M.}\ \bibnamefont
  {Denn}},\ }\bibfield  {title} {\enquote {\bibinfo {title} {{Pressure
  drop-flow rate equation for adiabatic capillary flow with a pressure- and
  temperature-dependent viscosity}},}\ }\href
  {https://doi.org/10.1002/pen.760210202} {\bibfield  {journal} {\bibinfo
  {journal} {Polymer Engineering and Science}\ }\textbf {\bibinfo {volume}
  {21}},\ \bibinfo {pages} {65--68} (\bibinfo {year} {1981})}\BibitemShut
  {NoStop}%
\bibitem [{\citenamefont {Schoof}(2007)}]{Schoof2007}%
  \BibitemOpen
  \bibfield  {author} {\bibinfo {author} {\bibfnamefont {C.}~\bibnamefont
  {Schoof}},\ }\bibfield  {title} {\enquote {\bibinfo {title}
  {{Pressure-dependent viscosity and interfacial instability in coupled
  ice–sediment flow}},}\ }\href {https://doi.org/10.1017/S0022112006002874}
  {\bibfield  {journal} {\bibinfo  {journal} {Journal of Fluid Mechanics}\
  }\textbf {\bibinfo {volume} {570}},\ \bibinfo {pages} {227} (\bibinfo {year}
  {2007})}\BibitemShut {NoStop}%
\bibitem [{\citenamefont {Srinivasan}, \citenamefont {Bonito},\ and\
  \citenamefont {Rajagopal}(2013)}]{Srinivasan2013}%
  \BibitemOpen
  \bibfield  {author} {\bibinfo {author} {\bibfnamefont {S.}~\bibnamefont
  {Srinivasan}}, \bibinfo {author} {\bibfnamefont {A.}~\bibnamefont {Bonito}},\
  and\ \bibinfo {author} {\bibfnamefont {K.~R.}\ \bibnamefont {Rajagopal}},\
  }\bibfield  {title} {\enquote {\bibinfo {title} {{Flow of a fluid through a
  porous solid due to high pressure gradients}},}\ }\href
  {https://doi.org/10.1615/JPorMedia.v16.i3.20} {\bibfield  {journal} {\bibinfo
   {journal} {Journal of Porous Media}\ }\textbf {\bibinfo {volume} {16}},\
  \bibinfo {pages} {193--203} (\bibinfo {year} {2013})}\BibitemShut {NoStop}%
\end{thebibliography}

%

\end{document}